\documentclass[ijoc,nonblindrev]{informs3} 

\OneAndAHalfSpacedXII 

\usepackage{natbib}
 \bibpunct[, ]{(}{)}{,}{a}{}{,}%
 \def\bibfont{\small}%

\definecolor{myblue}{RGB}{0, 20, 114}
\usepackage[hidelinks,colorlinks=true,urlcolor=myblue,linkcolor=myblue,citecolor=myblue]{hyperref}
\def\EMAIL#1{\href{mailto:#1}{#1}}

\def\theARTICLETOP{}
\def\theLRHFirstLine{}
\def\theLRHSecondLine{}
\def\theRRHFirstLine{}
\def\theRRHSecondLine{}

\usepackage{algorithm}
\usepackage{algpseudocode}
\usepackage{tikz}
\usepackage{braket}
\usepackage{booktabs}
\TheoremsNumberedThrough     

\EquationsNumberedThrough    

\begin{document}



\RUNTITLE{Quantum Grover Adaptive Search for Discrete Simulation Optimization}

\TITLE{Quantum Grover Adaptive Search for Discrete Simulation Optimization}

\ARTICLEAUTHORS{%
\AUTHOR{Mingjie Hu}
\AFF{School of Management, Fudan University \\
H. Milton Stewart School of Industrial and Systems Engineering, Georgia Institute of Technology\\
\EMAIL{23110690009@m.fudan.edu.cn}}
\AUTHOR{Jian-qiang Hu}
\AFF{School of Management, Fudan University, \EMAIL{hujq@fudan.edu.cn }}\AUTHOR{Enlu Zhou}
\AFF{H. Milton Stewart School of Industrial and Systems Engineering, Georgia Institute of Technology, \EMAIL{enlu.zhou@isye.gatech.edu}}
} 

\ABSTRACT{%
Quantum computing has advanced rapidly in recent years and has shown advantages in a variety of domains. In this paper, we investigate its potential for discrete simulation optimization in the fixed-confidence setting, a fundamental problem in the simulation literature. We first introduce a quantum simulation oracle that prepares a coherent superposition over all candidate solutions and provides the foundation for quantum algorithm design. Based on this oracle, we develop the first Grover-search-based quantum algorithm for discrete simulation optimization, called SOGAS. In particular, SOGAS uses a binary-search framework to progressively eliminate suboptimal solutions while carefully controlling the error probability, and eventually identifies a set of near-optimal solutions. We prove that SOGAS returns a near-optimal solution with probability at least the prescribed confidence level and achieves a quadratic speedup in the dependence of query complexity on the number of candidate solutions. Numerical experiments further show that SOGAS substantially outperforms classical benchmarks and provide empirical evidence for quantum advantage in discrete simulation optimization. 
}%



\maketitle

%


\section{Introduction}
Quantum computing has advanced rapidly in recent years and has shown promise in several areas of operations research, including combinatorial optimization \citep{abbas2024challenges}, reinforcement learning \citep{meyer2022survey}, and simulation \citep{blanchet2025quantum}. In simulation, previous work \citep{montanaro2015quantum, kothari2023mean} has shown that quantum algorithms can provide a quadratic speedup over classical Monte Carlo methods. In particular, to estimate the expectation of a random variable within a given additive error, a quantum algorithm needs only the square root of the number of oracle queries required by a classical method. This leads to a natural question in simulation optimization: can this quadratic speedup in estimation also be carried over to optimization? More specifically, can quantum simulation optimization algorithms provably reduce oracle query complexity by a quadratic factor compared with classical methods?

To address this question, we focus on a class of unstructured discrete simulation optimization problems known as ranking and selection (R\&S) in the simulation literature. In this setting, one is given a finite set of candidate solutions, each with stochastic performance represented by a random variable, and the goal is to identify the optimal solution in terms of expected performance through simulation experiments. In the fixed-confidence setting, the objective is to ensure that the probability of correct selection exceeds a prescribed threshold while minimizing the total number of simulation replications.

Some intuition can be gained by first considering a noiseless version of the discrete simulation optimization problem. Without stochastic noise, the problem reduces to an unstructured search problem. Classically, identifying the optimal solution among $N$ alternatives generally requires $O(N)$ operations, whereas Grover’s quantum search algorithm can find the optimum in only $O(\sqrt{N})$ operations. This gives a quadratic speedup, which is shown to be asymptotically optimal \citep{nielsen2010quantum}. These observations naturally lead to the question of whether Grover’s algorithm can be extended to discrete simulation optimization in the presence of stochastic noise, and whether such a quadratic speedup can still be achieved.

However, extending Grover’s algorithm to discrete simulation optimization presents several fundamental challenges:
\begin{itemize}
    \item \textbf{Quantum simulation system.} 
    A key challenge is to construct a quantum analogue of the classical simulation system that coherently encodes all candidate solutions in superposition. This is essential for quantum speedup, since it enables simultaneous operations on all solutions while faithfully representing the underlying stochastic simulation model.
    
    \item \textbf{Measurement-free design.} In quantum computing, measurement is the only way to extract information, but it collapses the quantum state and destroys the superposition needed for quantum parallelism. To preserve coherence, the algorithm must therefore avoid intermediate measurements. This makes standard classical approaches inapplicable.
    
    \item \textbf{Optimal-solution oracle construction.} Grover’s algorithm requires an oracle that can recognize the optimal solution. However, the optimal solution is unknown in advance, so such an oracle is not directly available. This prevents Grover’s algorithm from being applied directly.
    \item \textbf{Error control and performance guarantees.} In the fixed-confidence setting, the error probability must be carefully controlled so that the quantum algorithm returns a near-optimal solution with high probability. At the same time, one must establish that the algorithm achieves a quadratic speedup. This analysis is challenging, because classical proof techniques do not carry over directly to the quantum setting.
\end{itemize}

To help the reader, we interpret the main quantum concepts used in this paper through the lens of discrete simulation optimization. In particular, a quantum simulation system is represented by a quantum simulation oracle, which serves as the quantum analogue of a classical simulation model. In the classical setting, one simulation run for a given solution produces a single random performance observation. In the quantum setting, one query to the simulation oracle returns a quantum state that coherently encodes all candidate solutions and supports parallel operations across them. Measuring the state for a particular solution then yields a single random performance observation. Under this correspondence, the number of simulation replications in the classical framework becomes the number of quantum simulation oracle queries in the quantum framework, which we refer to as the \emph{query complexity}.

\subsection{Main Contributions}
The main contribution of this paper are summarized as follows:
\begin{itemize}
    \item We study discrete simulation optimization in the fixed-confidence setting. To achieve quantum speedup, we introduce a quantum simulation oracle that prepares a coherent superposition over all candidate solutions, enabling parallel search across the solution space.
    
    \item We develop a Grover Adaptive Search (SOGAS) algorithm, for discrete simulation optimization. The algorithm avoids frequent intermediate measurements and preserves quantum coherence, which is critical for realizing quantum speedup. Specifically, SOGAS uses a binary-search-based framework to progressively eliminate suboptimal solutions, identifies a near-optimal set of solutions, and then applies amplitude amplification to output one near-optimal solution.
    
    \item We establish theoretical performance guarantees for SOGAS. In particular, we prove that the algorithm returns a near-optimal solution with at least the prescribed confidence level. We further show that its query complexity achieves a quadratic speedup, in its dependence on the number of candidate solutions, over classical algorithms. 
    
    \item Finally, we provide numerical results to validate the quantum advantage and offer additional insight, which we believe is especially valuable given that most existing work on quantum algorithms is purely theoretical.
    We implement SOGAS in Qiskit-based simulations, and the numerical results demonstrate a substantial quantum advantage in query complexity over its classical counterpart.
\end{itemize}

To the best of our knowledge, this is the first work in the discrete simulation optimization literature to use quantum computing to establish a quadratic improvement in query complexity with respect to the number of solutions. Our results provide a positive answer to the question raised earlier: the quadratic speedup achieved in estimation can indeed be extended to optimization. More broadly, the proposed problem formulation and algorithmic framework open a new avenue for future research on exploiting quantum advantages in classical Monte Carlo simulation and simulation optimization.

\subsection{Related Literature}

R\&S, a major class of discrete simulation optimization, is one of the most extensively studied problems in simulation. Existing formulations generally fall into two main categories. In the fixed-budget setting, the goal is to maximize the probability of correctly identifying the optimal solution subject to a given simulation budget \citep{chen2000simulation,glynn2004large,peng2018ranking}. In the fixed-confidence setting, the goal is to ensure that the probability of incorrectly selecting a non-optimal solution remains below a prescribed risk level \citep{kim2001fully,fan2016indifference}. More recent research has considered variant formulations that incorporate contextual information \citep{shen2021ranking,du2024contextual,li2024efficient}, stochastic constraints \citep{lee2012approximate,hu2024multi}, structural information \citep{hu2025optimal}, and input uncertainty \citep{wu2024data}.

There is existing work on extending Grover’s algorithm to deterministic optimization problems \citep{bulger2003implementing,baritompa2005grover,gilliam2021grover}. In contrast, we study a simulation optimization problem in which the objective function can only be evaluated through noisy simulation. As a result, the exact comparisons required by those deterministic methods cannot be implemented directly, rendering them inapplicable in our setting.

Our work is inspired by recent work in simulation that explores the use of quantum computing to improve simulation optimization \citep{ zabinsky2025comprehensive}. For example, \citet{gacon2020quantum} proposed a quantum-enhanced simulation-based optimization framework that mainly uses quantum amplitude estimation to accelerate stochastic objective evaluation within a classical optimization loop. In contrast, our work leverages Grover-style search to accelerate the discrete search process in simulation optimization. Moreover, their approach is largely heuristic and does not provide theoretical guarantees. However, we study the problem in the fixed-confidence setting and establish rigorous query complexity guarantees that demonstrate a quantum advantage over classical methods.

To the best of our knowledge, the most closely related work is \cite{hu2025quantum}, which studies discrete simulation optimization with stochastic constraints. However, our work differs from it in three fundamental ways. First, the quantum simulation oracle is different: our oracle prepares a coherent superposition over all solutions, which enables a stronger quantum speedup. Second, our query complexity analysis focuses on its overall dependence on the total number of solutions. In contrast, under the quantum simulation oracle considered in \cite{hu2025quantum}, the query complexity depends linearly on the number of solutions, and the quadratic speedup applies only to the evaluation of each solution. Moreover, because the underlying quantum simulation oracles differ, their algorithmic framework cannot be directly transferred to our setting.

Recent studies have also explored quantum algorithms for best-arm identification (BAI) \citep{casale2020quantum,wang2021quantum,wang2025quantum}, a core machine learning problem closely related to discrete simulation optimization. However, our work differs in its problem formulation, algorithm design, and optimality guarantees. In particular, \cite{wang2021quantum} studies the BAI problem and develops a quantum algorithm for identifying the unique best arm by conceptually extending Grover’s search. In contrast, we consider a simulation optimization problem and aim to identify an $\epsilon$-optimal solution rather than the exact optimum. Our goal is to achieve a quadratic speedup in query complexity with respect to the number of candidate solutions, whereas their objective is to match the exact quantum query complexity lower bound for BAI. In addition, we assume general bounded random performance, while their analysis focuses primarily on Bernoulli rewards. These differences in formulation and optimality criteria lead to a simpler algorithmic design and different theoretical guarantees in our setting. 

The remainder of this paper is organized as follows. Section \ref{sec: preliminaries} reviews preliminaries on quantum computing and Grover’s algorithm. Section \ref{sec: formulation} formulates the discrete simulation optimization problem, defines the quantum simulation oracle, and presents the Grover Adaptive Search algorithm. Section \ref{sec: analysis} shows that the proposed algorithm identifies a near-optimal solution with high probability and achieves a quadratic speedup. Section \ref{sec: experiment} presents numerical experiments to illustrate the empirical performance of the algorithm. Finally, Section \ref{conclusion} concludes the paper and discusses directions for future research. Technical proofs are deferred to the appendix.

\section{Preliminaries}
\label{sec: preliminaries}
In this section, we briefly review the basic concepts and preliminaries of quantum computing. We then introduce the framework of Grover’s algorithm, which forms the foundation of our algorithm design. For a comprehensive treatment, see \cite{nielsen2010quantum}.

\subsection{Preliminaries on Quantum Computing}
Quantum computing exploits phenomena such as superposition, entanglement, and interference to process information in ways that can significantly accelerate certain computations. By preparing a quantum state in superposition over candidate solutions and applying a unitary transformation that encodes their stochastic performance, one obtains a state that captures the performance information of all solutions simultaneously. This form of quantum parallelism is a key source of potential speedup.

In quantum computing, Dirac’s bra-ket notation is used to represent vectors in complex Hilbert spaces. A ket $|\psi\rangle = (\psi_1, \dots, \psi_n)^\top \in \mathbb{C}^n$ denotes a column vector, and its corresponding bra $\langle\psi| = (\psi_1^\star, \dots, \psi_n^\star)$ is its conjugate transpose. For any two states $|\psi\rangle, |\phi\rangle \in \mathbb{C}^n$, the inner product is defined as $$\langle\psi|\phi\rangle := \sum_{i=1}^n \psi_i^\star \phi_i \in \mathbb{C},$$ which induces the $\ell_2$-norm $\|\,|\psi\rangle\|_2 = \sqrt{\langle\psi|\psi\rangle}$. The outer product of $\ket{\psi}$ and $\ket{\phi}$ is defined as
\[
|\psi\rangle\langle\phi| \in \mathbb{C}^{n\times n},
\]
whose $(i,j)$-th entry is $\psi_i \phi_j^\star$. The tensor product of two quantum states $|\psi\rangle \in \mathbb{C}^{n_1}$ and $|\phi\rangle \in \mathbb{C}^{n_2}$ is defined as \[
|\psi\rangle|\phi\rangle
:=
|\psi\rangle \otimes |\phi\rangle
=
(\psi_1\phi_1,\psi_1\phi_2,\dots,\psi_1\phi_{n_2},\psi_2\phi_1,\dots,\psi_{n_1}\phi_{n_2})^\top
\in
\mathbb{C}^{n_1}\otimes\mathbb{C}^{n_2},
\] which represents the joint state of the composite system.

The basic unit of information in a quantum computer is the qubit. Unlike a classical bit, which can take only one of the two binary states $0$ or $1$, a qubit can exist in a linear combination of these states, known as a quantum superposition. Mathematically, a single-qubit state is represented as $|\psi\rangle = \alpha|0\rangle + \beta|1\rangle$, where $\alpha, \beta \in \mathbb{C}$ are complex probability amplitudes. These amplitudes must satisfy the normalization condition $|\alpha|^2 + |\beta|^2 = 1$. Upon measurement in the computational basis, the state collapses to $|0\rangle$ with probability $|\alpha|^2$ or $|1\rangle$ with probability $|\beta|^2$.

\subsection{Grover's Algorithm}

In this subsection, we briefly review Grover's algorithm, a cornerstone quantum search algorithm for locating a target item in an unstructured database. The primary advantage of Grover's algorithm lies in its ability to significantly accelerate the search process. In classical computing, identifying a target entry within a set of $N$ unsorted elements requires $O(N)$ operations in the worst-case scenario. In contrast, Grover's algorithm accomplishes this task using only $O(\sqrt{N})$ quantum operations and achieves a quadratic speedup over any classical counterpart.

There are three basic components in Grover’s search algorithm:
\begin{enumerate}
    \item \textbf{State initialization.}
    We use $n$ qubits initialized at $\ket{0}_{n}$ to encode a search space of size $N=2^n$.
    Applying Hadamard gates $H^{\otimes n}$ to all qubits prepares the uniform superposition state:
\begin{equation} H^{\otimes n} \ket{0}_n = \frac{1}{\sqrt{N}}\sum_{x=0}^{N-1}\ket{x}. \end{equation}

    \item \textbf{Phase oracle.}
    Let $\mathcal{M} \subseteq \{0,1,\ldots,N-1\}$ denote the set of labeled (target) indices.
    The oracle ${O}_{f}$ is a phase-flip operator that recognizes the labeled states and flips their phases.
    Consequently, after one oracle call, the superposition becomes
    \begin{equation}
        {O}_{f}H^{\otimes n} \ket{0}_n
        = \frac{1}{\sqrt{N}}\sum_{x\notin \mathcal{M}}\ket{x}
        - \frac{1}{\sqrt{N}}\sum_{x\in \mathcal{M}}\ket{x}.
    \end{equation}

    \item \textbf{Diffusion operator.}
    Grover's diffusion operator reflects the state about the uniform superposition, defined as
    $
        D := H^{\otimes n}\left(2\ket{0}\bra{0} - I\right)H^{\otimes n},
    $
    where $I$ is the identity operator.  Since $H^{\otimes n}\ket{0}$ is the uniform superposition state, the operator $D$ implements a reflection about this state,  
    which is often interpreted as an inversion about the mean of amplitudes.
\end{enumerate}

The Grover iteration is defined as
$ G := D {O}_{f},$
and the algorithm repeatedly applies $G$ to the initial superposition state $H^{\otimes n}\ket{0}^{\otimes n}$.
After an appropriate number of iterations, the amplitudes of the labeled states are amplified (and those of unlabeled states are suppressed), yielding a high probability of measuring an element of $\mathcal{M}$.

\begin{figure}
\centering
\includegraphics[width=0.9\linewidth]{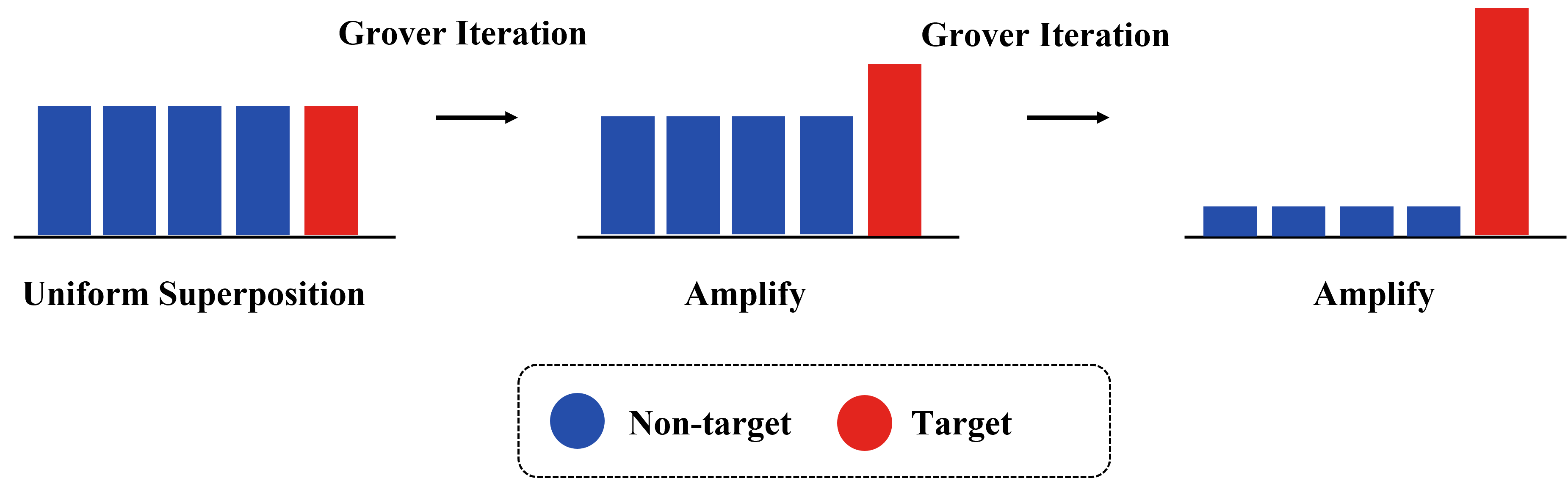}
\caption{Grover's Algorithm}
\label{fig:grover}
\end{figure}

Figure \ref{fig:grover} illustrates the basic idea of Grover’s iteration. The algorithm begins with an equal superposition over all items in the search space, so each item initially has the same amplitude. It then repeatedly applies the Grover iteration, which flips the phase of the basis state corresponding to the labeled item and reflects the resulting state about the average amplitude. As a result, the amplitude of the labeled item increases over time, while the amplitudes of the unlabeled items decrease relative to it. After a sufficient number of iterations, most of the probability mass is concentrated on the labeled item, so a measurement returns it with high probability.

\section{Quantum Discrete Simulation Optimization}
\label{sec: formulation}
In this section, we first formulate the discrete simulation optimization problem in the fixed-confidence setting. We then define the quantum simulation oracle and briefly discuss how it can be implemented in practice. Finally, we develop a Grover-search-based algorithm that extends Grover’s search framework to the more challenging setting of stochastic simulation optimization.

\subsection{Discrete Simulation Optimization}
We consider a stochastic system whose performance depends on a discrete decision variable $x$ selected from a finite set $\mathcal{X}$. For each $x\in\mathcal{X}$, the realized system performance is determined by two components: a random input $\xi_x$, defined on a support set $\Xi$ with distribution $P(\xi_x)$, and a deterministic performance function $Y:\mathcal{X}\times \Xi\rightarrow \mathbb{R}$, which maps $(x,\xi_x)$ to a realized performance value. The random input $\xi_{x}$ captures all uncertainty in the simulation under decision $x$. In general, the function $Y$ is induced by the underlying simulation dynamics and typically does not admit a closed-form expression. We define the objective value of decision $x$ as its mean performance,
\begin{equation*}
    y(x) := \mathbb{E}[Y(x,\xi_x)].
\end{equation*}

The value of $y(x)$ cannot be computed exactly and must instead be estimated from simulation output. Let $x^\star$ denote an optimal solution, defined by
\begin{equation*}
    x^\star \in \argmax_{x\in \mathcal{X}} y(x).
\end{equation*}
Our goal is to develop a simulation optimization algorithm that identifies a near-optimal solution $x$ satisfying
\begin{equation}
\label{eq: epsilon_guarantee}
    y(x^\star) - y(x) \le \epsilon,
\end{equation}
where $\epsilon>0$ is a prescribed tolerance parameter, also referred to as the optimality gap. Since $y(x)$ is the expectation of a stochastic simulation output and is not available in closed form, we impose no structural assumptions on the objective function. Moreover, due to simulation noise, the guarantee in \eqref{eq: epsilon_guarantee} generally cannot be achieved with probability one under a finite simulation budget. We therefore adopt the standard $(\epsilon,\delta)$-probably approximate correct ($(\epsilon,\delta)$-PAC, see Definition \ref{def: delta-PAC}) criterion.
\begin{definition}[$(\epsilon,\delta)$-PAC algorithm]
\label{def: delta-PAC}
For given parameters $\epsilon>0$ and $\delta\in(0,1)$, a simulation optimization algorithm is called $(\epsilon,\delta)$-PAC if it returns a solution $x$ such that
\[
\mathbb{P}\bigl(y(x^\star)-y(x)>\epsilon\bigr)\le \delta.
\]
\end{definition}

To facilitate algorithm design and establish the $(\epsilon,\delta)$-PAC guarantee for this general setting, we impose the following assumption on the simulation output.
\begin{assumption}
\label{ass: boundness}
For every $x\in\mathcal{X}$, the random performance $Y(x,\xi_x)$ is bounded in $[0,1]$.
\end{assumption}

\begin{remark}
Without loss of generality, Assumption \ref{ass: boundness} rescales any bounded performance distribution to the interval $[0,1]$. For an unbounded distribution, such as a Gaussian distribution, one may truncate the support to a bounded interval. This is also the approach used in our Qiskit-based quantum simulations to handle continuous random variables. As a consequence of Assumption \ref{ass: boundness}, the expected performance $y(x)$ is also bounded in $[0,1]$.
\end{remark}

\subsection{Quantum Simulation Oracle}
In this subsection, we formalize the notion of an oracle for discrete simulation optimization by first introducing its classical counterpart and then extending it to the quantum setting.

In classical simulation optimization, the stochastic system can be viewed as a random oracle. Running the simulation model under a decision variable $x$ is equivalent to querying the corresponding oracle, and each query returns an independent random observation of $Y(x,\xi_x)$. 

To define the quantum simulation oracle, we use three registers: the solution register $\mathcal X$, the noise register $\xi$, and the output register $Y$. The state $\ket{\omega}_{\mathcal X}\ket{0}_{\xi}\ket{0}_{Y}$ represents an input in which the solution register is prepared in the superposition state $\ket{\omega}_{\mathcal X}$, while the noise and output registers are respectively initialized in the zero state.

To design quantum algorithms, we introduce a quantum analogue of the classical simulation system, called the quantum simulation oracle, in Definition \ref{def: quantum oracle}. Candidate solutions are accessed in superposition by applying the unitary oracle $\mathcal Q$ to the state $\ket{\omega}_{\mathcal X}\ket{0}_{\xi}\ket{0}_{Y}$. A classical query to solution $x$ corresponds to choosing $\ket{\omega}_{\mathcal X}=\ket{x}_{\mathcal X}$ and measuring the output register, which produces a random observation of $Y(x,\xi_x)$. Therefore, the number of simulation replications in the classical setting corresponds naturally to the number of oracle queries in the quantum setting.

In the quantum simulation oracle, the input register $\mathcal{X}$ is prepared in the superposition state
\[|\omega\rangle=\sum_{x\in\mathcal X}\omega_x|x\rangle,\]
where $\sum_{x\in\mathcal X}|\omega_x|^2=1$. The oracle $\mathcal Q$ acts coherently on all branches of this superposition, and produces a joint quantum state that encodes, in superposition, both the random noise $\xi_x$ distributed according to $P(\xi_x)$ and the corresponding simulation output $Y(x,\xi_x)$, as specified in the following definition.

\begin{definition}[Quantum Simulation Oracle]
\label{def: quantum oracle}
A quantum simulation oracle is a unitary operator $\mathcal Q$ such that
\begin{equation}
\label{eq: simulation oracle}
\mathcal Q\; \ket{\omega}_{\mathcal{X}}\ket{0}_{\xi}\ket{0}_{Y}
\;\mapsto\;
\sum_{x\in\mathcal X}\omega_x\ket{x}_{\mathcal{X}}
\left(\sum_{\xi_x\in\Xi}\sqrt{P(\xi_x)}\,\ket{\xi_x}_{\xi}\ket{Y(x,\xi_x)}_{Y}\right),
\end{equation}
where $\Xi$ is the support of the random noise $\xi_x$, and $P(\xi_x)$ is its probability mass function. We allow $\mathcal Q$ to use additional ancilla or workspace registers to ensure reversibility \citep{nielsen2010quantum}; these are omitted for simplicity.
\end{definition}
\begin{remark}
For simplicity, the quantum simulation oracle in \eqref{eq: simulation oracle} is defined for discrete random variables, which is also common in the related literature \citep{gacon2020quantum, dai2023quantum,hu2025quantum}. Continuous distributions, such as Gaussian distributions, can be approximated arbitrarily well by discretizing their support. This is also the approach adopted in our Qiskit-based simulation.
\end{remark}

A central question is how to implement the simulation oracle in the quantum setting. Following \citet{wang2021quantum2}, when the simulator is given by a classical computer program with accessible source code, it can be converted into a quantum circuit with only polynomial overhead. The resulting circuit can then be executed coherently on superpositions of inputs and used as a quantum simulation oracle. This observation supports the quantum simulation model adopted in our work. In addition, several studies have investigated the construction of quantum simulation models for specific applications, such as queueing systems and reinforcement learning; see, for example, \citet{esa2024single} and \citet{dunjko2016quantum}. Therefore, in this paper, we assume that such a quantum simulation model is available and fixed, and focus on developing simulation optimization algorithms based on such a quantum simulation model.

\subsection
{Grover Adaptive Search}
In this subsection, we develop a Grover-search-based algorithm for simulation optimization. The overall framework, given in Algorithm~\ref{alg:grover} (SOGAS), adopts the main idea of Grover’s search algorithm. SOGAS first constructs the subroutine \textsc{OptimalRegion}, presented in Algorithm~\ref{alg:confidence_interval}, to identify a region $\mathcal{R}$ that contains optimal mean performance. Based on this region, SOGAS then builds the quantum flagging algorithm $\mathcal{A}$, presented in Algorithm~\ref{alg:qbso}, which serves as an oracle for labeling near-optimal solutions. Next, SOGAS applies the standard amplitude amplification technique \citep{brassard2000quantum} to increase the amplitudes of the labeled solutions. Finally, measuring the amplified state returns a near-optimal solution with high probability. 

Amplitude amplification generalizes Grover’s search algorithm to an arbitrary quantum subroutine by amplifying the initial success probability of the subroutine. Grover’s algorithm is a special case of this framework in which the initial state is the uniform superposition over the search space. This framework provides a quadratic speedup for a broad class of search problems. Lemma~\ref{lem: amplitude amplify} states the query complexity required for amplitude amplification to output a labeled near-optimal solution.

\begin{lemma}[Amplitude amplification in \citealt{brassard2000quantum}] 
\label{lem: amplitude amplify}
Let $\mathcal B$ be a quantum algorithm with query complexity $T_{\mathcal B}$, and let $p_{\mathcal B}$ denote the probability that measuring the output state of $\mathcal B$ yields a labeled solution. Then there exists a quantum algorithm $\mathrm{Amplify}(\mathcal B,\tilde\delta)$ that uses
\[
O\left(
\frac{T_{\mathcal B}}{\sqrt{p_{\mathcal B}}}
\log\frac{1}{\tilde\delta}
\right)
\]
queries to the underlying oracle and its inverse, and outputs a labeled solution with probability at least $1-\tilde\delta$. 
\end{lemma}

\begin{algorithm}[t]
\caption{Simulation Optimization via Grover Adaptive Search (SOGAS)}
\label{alg:grover}
\begin{algorithmic}[1]
\Require Quantum simulation oracle $\mathcal{Q}$, optimality gap $\epsilon \in (0,1)$, risk level $\delta \in (0,1)$
\State $\mathcal{R} \gets \textsc{OptimalRegion}(\mathcal{Q}, \epsilon, \delta/2)$ \Comment{Construct a region containing $y(x^\star)$}
\State $a \gets \min \mathcal{R}$, $b \gets \max \mathcal{R}$
\State $\eta \gets (b-a)/4$, $\ell \gets a-\frac{3}{4}\eta$
\State $\mathcal{A}_{\mathrm{flag}} \gets \textsc{FlagOracle}(\mathcal{Q}, \ell, \eta, 0.01\delta)$ \Comment{Label near-optimal solutions}
\State $\hat{x}^\star \gets \textsc{Amplify}(\mathcal{A}_{\mathrm{flag}}, \delta/2)$\Comment{Amplify the amplitudes of target solutions}
\State \Return $\hat{x}^\star$
\end{algorithmic}
\end{algorithm}

The central challenges in developing the SOGAS algorithm are twofold. First, unlike in the standard Grover search setting, we do not have direct access to an oracle that labels the target solutions. Second, because the simulation outputs are stochastic, the algorithm must control the error probability carefully to ensure that the returned solution is $\epsilon$-optimal with probability at least $1-\delta$. 

\begin{figure}
    \centering
    \includegraphics[width=0.8\linewidth]{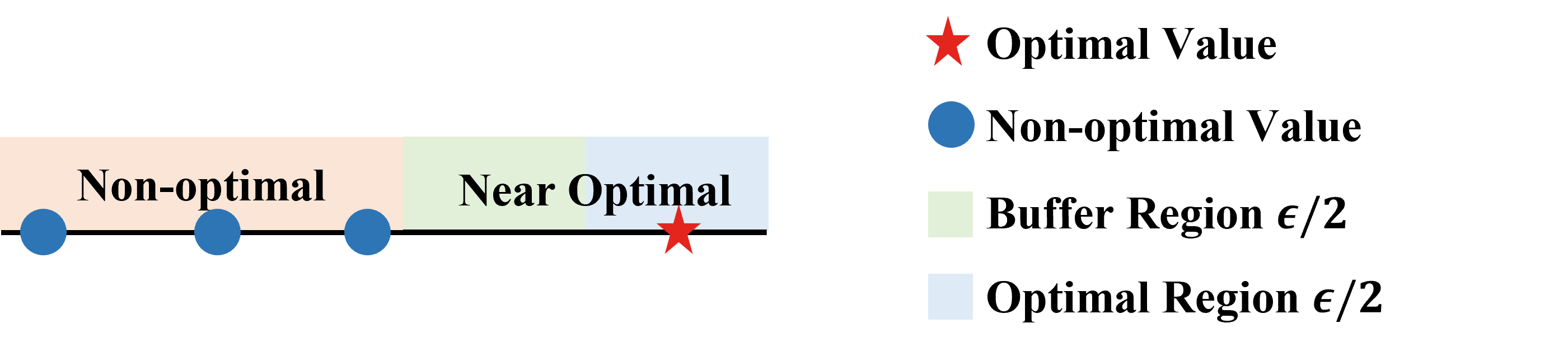}
    \caption{Optimal Region in Algorithm \ref{alg:grover}}
    \label{fig:opt_region}
\end{figure}

The main idea for addressing these challenges is illustrated in Figure~\ref{fig:opt_region}. We partition the performance interval into three disjoint subregions: the optimal region $\mathcal{R}$, a buffer region, and a non-optimal region. The optimal region is constructed to contain the optimal mean performance $y(x^\star)$ with high probability and to have width less than $\epsilon/2$. The buffer region is introduced to account for the uncertainty from quantum estimation and is also chosen to have a width less than $\epsilon/2$. The rest of the interval is classified as non-optimal. Solutions whose mean performances fall in the optimal region, as well as some solutions in the buffer region that are indistinguishable from them under the quantum estimation error, are labeled solutions.

Therefore, if the optimal mean performance $y(x^\star)$ lies in $\mathcal{R}$ and the combined width of the optimal and buffer regions is less than $\epsilon$, then all labeled solutions are $\epsilon$-optimal. Amplitude amplification then returns one such labeled solution with high probability.

\begin{figure}
\centering
\includegraphics[width=0.8\linewidth]{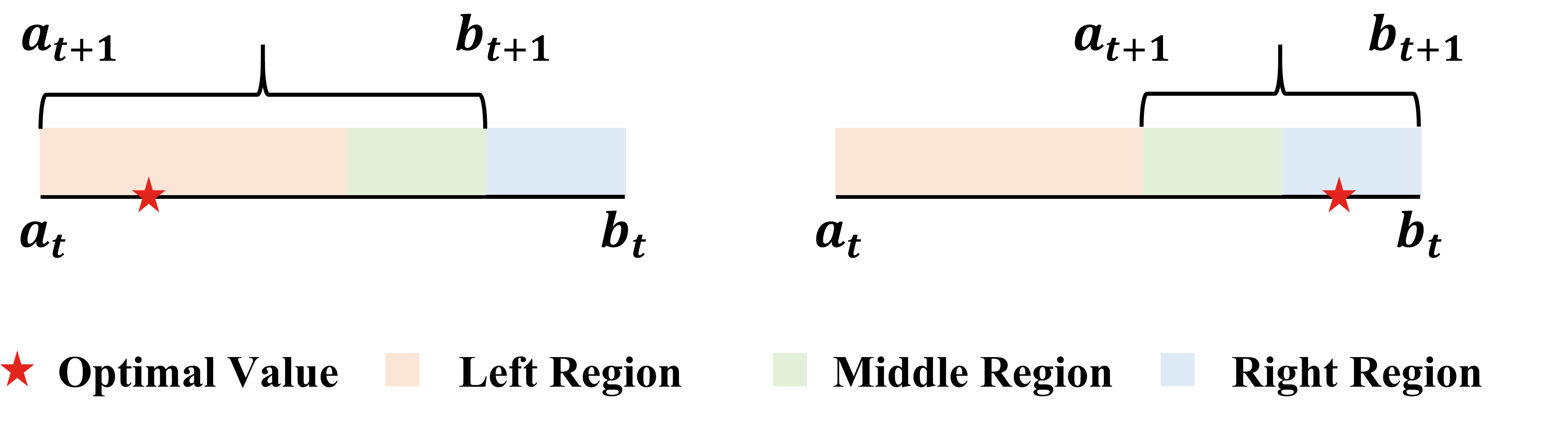}
\caption{Binary Search Process in Algorithm \ref{alg:confidence_interval}}
\label{fig:binary_search}
\end{figure}

To implement this idea, we first construct an optimal region $\mathcal{R}$ that contains the optimal performance $y(x^\star)$ with high probability. This is achieved through a binary-search-based procedure, illustrated in Figure~\ref{fig:binary_search} and formally described in Algorithm~\ref{alg:confidence_interval}. At iteration $t$, the current interval $[a_t,b_t]$ is divided into three disjoint subregions. The middle subregion serves as a buffer to account for the uncertainty introduced by the quantum estimation procedure, while the left and right subregions are the two candidate parts to be either retained or discarded. If the optimal solution lies in the middle or right subregion, then the left subregion can be discarded. Conversely, if it lies in the middle or left subregion, then the right subregion can be discarded. Repeating this procedure progressively shrinks the interval until its width is less than $\epsilon/2$. Throughout the process, the error probability at each iteration is carefully controlled so that the probability that the constructed region $\mathcal{R}$ fails to contain the optimal performance $y(x^\star)$ remains within the prescribed tolerance. 

\begin{algorithm}[t]
\caption{\textsc{OptimalRegion} ($\mathcal{Q}$, $\epsilon$, $\delta$)}
\label{alg:confidence_interval}
\begin{algorithmic}[1]
\Require Quantum simulation oracle $\mathcal{Q}$, optimality gap $\epsilon \in (0,1)$, risk level $\delta \in (0,1)$, 
\State $\mathcal{R}_0 \gets [0,1]$
\State $t \gets 0$
\While{$|\mathcal{R}_t| \ge \epsilon/2$}
    \State $a_t \gets \min \mathcal{R}_t$, $b_t \gets \max \mathcal{R}_t$
    \State $\eta_t \gets (b_t-a_t)/4$, $\ell_t \gets (a_t+b_t)/2$
    \State $\delta_t \gets \delta / 2^{t+1}$
    \State Augment $\mathcal{Q}$ with an auxiliary solution $x_a$ satisfying $y(x_a)=1$, and denote the resulting oracle by $\bar{\mathcal{Q}}_t$ \Comment{Establish an upper bound on the performance}
    \State Construct the quantum flag algorithm $\mathcal{A}_t \gets \textsc{FlagOracle}(\bar{\mathcal{Q}}_t,\ell_t,\eta_t,0.1\delta_t)$
    \State $r_t \gets \textsc{Estimate}(\mathcal{A}_t,0.1,\delta_t)$ \Comment{Estimate the proportion of labeled solutions; see Lemma \ref{lemma: ae}}
    \State $m_t \gets \left\lceil \log_2 \frac{1}{2\eta_t} \right\rceil + 2$
    \If{$r_t > \frac{3}{2(|\mathcal{X}|+1)}$}
        \State $\mathcal{R}_{t+1} \gets [\,\ell_t+\eta_t-2^{1-m_t},\, b_t\,]$
    \Else
        \State $\mathcal{R}_{t+1} \gets [\,a_t,\, \ell_t+\eta_t-2^{-m_t}\,]$
    \EndIf
    \State $t \gets t+1$
\EndWhile
\State \Return $\mathcal{R}_t$
\end{algorithmic}
\end{algorithm}

Algorithm~\ref{alg:confidence_interval} applies the standard amplitude estimation technique within the binary-search procedure to estimate the proportion of solutions whose performances exceed a deterministic threshold $r_t$. Lemma~\ref{lemma: ae} gives the query complexity required for amplitude estimation to achieve a prescribed estimation precision.

\begin{lemma}[Theorem 6 in \citealt{wang2021quantum}]
\label{lemma: ae}
Let $O_r$ be a unitary such that $$O_r\ket{0}=\sqrt{1-r}\ket{0}+\sqrt{r}\ket{1}.$$ Then, for any $\tilde\epsilon,\tilde\delta\in(0,1)$, there exists a quantum algorithm $\textsc{Estimate}(O_r,\tilde\epsilon,\tilde\delta)$ that uses $O(\frac{1}{\tilde\epsilon}\log(\frac{1}{\tilde\delta}))$ queries to $O_r$ and its inverse and outputs an estimate $\hat r$ satisfying \[\mathbb{P}(|\hat r-r|\le \tilde\epsilon)\ge 1-\tilde\delta.\]
\end{lemma}

\begin{algorithm}[t]
\caption{\textsc{FlagOracle} ($\mathcal{Q}$, $\ell$, $\eta$, $\kappa$)}
\label{alg:qbso}
\begin{algorithmic}[1]
\Require Quantum simulation oracle $\mathcal{Q}$, reference point $\ell$, precision $\eta$, risk level $\kappa$
\State $\Delta \gets 2\eta$
\State $m \gets \left\lceil \log_2 \frac{1}{\Delta} \right\rceil + 2$
\State $\alpha \gets \frac{\kappa^2}{4|\mathcal{X}|^{3}}$
\State Prepare the state
\Statex
\begin{equation}
\label{eq: superposition}
\frac{1}{\sqrt{|\mathcal{X}|}}
\sum_{x\in\mathcal{X}} \ket{x}_{\mathcal{X}}
\sum_{\xi_x\in\Xi}\sqrt{P(\xi_x)}\,\ket{\xi_x}_{\xi}\ket{Y(x,\xi_x)}_{Y}
\ket{0}_{A}\ket{0}_{P}\ket{0}_{F}
\end{equation}
\State For each $x\in\mathcal{X}$, conditioned on register $\mathcal{X}$ being in state $\ket{x}_{\mathcal{X}}$, apply $\textsc{QAE}(2^{-m}/3,\alpha)$ to oracle $\mathcal{Q}_x$, where
\[
\mathcal{Q}_x\ket{0}_{\xi}\ket{0}_{Y}
=
\sum_{\xi_x\in\Xi}\sqrt{P(\xi_x)}\,\ket{\xi_x}_{\xi}\ket{Y(x,\xi_x)}_{Y}
\] \Comment{QAE is a quantum mean estimation algorithm; see Lemma \ref{lemma: qae}}
\State Coherently compare the output estimate in the register $P$ with the midpoint of the interval
$[\ell+\eta-2^{1-m},\,\ell+\eta-2^{-m}]$, and store the resulting label in register $A$
\Comment{see Lemma \ref{lem: gapped estimation}}
\State Apply a controlled-NOT gate with control register $A$ and target register $F$
\State \Return the resulting quantum procedure
\end{algorithmic}
\end{algorithm}

We develop Algorithm~\ref{alg:qbso} to partition the mean performance interval into three disjoint subregions and label the target solutions accordingly. The algorithm starts from the uniform superposition state defined in \eqref{eq: superposition}. It uses three additional registers: register $A$ stores the outcome of the threshold comparison for the mean performance of each solution, register $P$ stores the output of the quantum mean estimation subroutine, and register $F$ serves as the final flag register.

For each $x \in \mathcal{X}$, the algorithm applies the quantum mean estimation procedure $\textsc{Qae}(\epsilon,\delta)$ to estimate $\mathbb{E}[Y(x,\xi_x)]$. If the estimate is close to the right endpoint of the interval $[\ell+\eta-2^{1-m},\ell+\eta-2^{-m}]$, then the algorithm writes $1$ into register $A$; otherwise, it writes $0$. Therefore, register $A$ indicates whether the corresponding solution has sufficiently good performance. Next, a controlled-NOT gate is applied to transfer the value in register $A$ to register $F$. Consequently, solutions with sufficiently good performance are labeled by setting register $F$ to $1$. The entire procedure is implemented coherently, without intermediate measurement, so that the superposition over candidate solutions is preserved throughout.

Step 5 of Algorithm~\ref{alg:qbso} employs the quantum Monte Carlo mean estimation method proposed by \citet{kothari2023mean}. Lemma~\ref{lemma: qae} states the query complexity required for this procedure to achieve the desired probabilistic guarantee on the estimate.

\begin{lemma}[Theorem 1.1 in \citealt{kothari2023mean}]
\label{lemma: qae}
Let $Z$ be a random variable with mean $\mu$ and standard deviation $\sigma$, and suppose we have access to an oracle that outputs $Z$. Then there exists a quantum algorithm, denoted by $\textsc{QAE}(\tilde \epsilon, \tilde\delta)$, that makes
$
O(\frac{\sigma}{\tilde\epsilon}\log\frac{1}{\tilde\delta})
$
queries to this oracle and outputs an estimate $\hat{\mu}$ of $\mu$ such that
$
\mathbb{P}(|\hat{\mu}-\mu|\le \tilde\epsilon)\ge 1-\tilde\delta.
$
\end{lemma}
\begin{remark}
To obtain the same accuracy and confidence guarantee for classical Monte Carlo mean estimation, one needs
$
O(\frac{\sigma^2}{\tilde\epsilon^2}\log\frac{1}{\tilde\delta})
$
queries to the classical oracle, or in other words, the same number of samples. Thus, the quantum algorithm achieves a quadratic speedup in mean estimation.
\end{remark}

\section{Correctness and Query Complexity Analysis}
\label{sec: analysis}
In this section, we establish the statistical correctness of the SOGAS algorithm by proving its $(\epsilon,\delta)$-PAC guarantee, and we derive an upper bound on its query complexity to show that it achieves a quadratic speedup for discrete simulation optimization.

We begin by analyzing the quantum flagging procedure defined in Algorithm~\ref{alg:qbso}. 
Lemma~\ref{lem: gapped estimation} characterizes the output state after Steps~5 and~6 of Algorithm~\ref{alg:qbso}. The two thresholds, $\ell+\eta-2^{1-m}$ and $\ell+\eta-2^{-m}$, partition the entire interval into three disjoint subintervals. Due to the uncertainty in quantum mean estimation, for solutions whose true mean performances lie in the middle region, the value stored in register $A$ may be either $1$ or $0$. Consequently, all solutions in the right region, together with some proportion of the solutions in the middle region, will be labeled. The proof of this lemma relies on the probabilistic guarantee established in Lemma~\ref{lemma: qae}.
\begin{lemma}[Output State After Steps 5-6 of Algorithm \ref{alg:qbso}]
\label{lem: gapped estimation}
Let $\mathcal A=\mathcal A(\mathcal Q,\ell,\eta,\kappa)$ denote Algorithm~\ref{alg:qbso} with input parameters $\mathcal Q$, $\ell$, $\eta$, and $\kappa$. The output state after steps $5$ and $6$ of Algorithm \ref{alg:qbso} is 
\[
\frac{1}{\sqrt{|\mathcal X|}}
\sum_{x\in\mathcal X}
\ket{x}_{\mathcal X}
\ket{\mathrm{data}_x}_{\xi,Y}
\Bigl(
\lambda_{x,0}\ket{0}_A\ket{\gamma_{x,0}}_{P}
+
\lambda_{x,1}\ket{1}_A\ket{\gamma_{x,1}}_{P}
\Bigr)\ket{0}_{F},
\]
where for every $x\in\mathcal X$,
$
\lambda_{x,0},\lambda_{x,1}\in[0,1],
\lambda_{x,0}^2+\lambda_{x,1}^2=1,
$ and moreover, $\lambda_{x,0}^2\le \alpha$ whenever $\mathbb{E}[Y(x,\xi_x)] \ge \ell+\eta-2^{-m}$, and $\lambda_{x,1}^2\le \alpha$ whenever $\mathbb{E}[Y(x,\xi_x)] < \ell+\eta-2^{1-m}$.
\end{lemma}

Ideally, Algorithm \ref{alg:qbso} partitions the solution set $\mathcal{X}$ into three disjoint subsets, $\mathcal{S}_1,\mathcal{S}_2$ and $\mathcal{S}_3$ according to their mean performances:
\[
\mathcal{S}_{\gamma}
:=
\begin{cases}
\left\{x\in\mathcal{X}: y(x)\ge l+\eta-2^{-m}\right\}, & \gamma=1,\\[2mm]
\left\{x\in\mathcal{X}: y(x)< l+\eta-2^{1-m}\right\}, & \gamma=2,\\[2mm]
\left\{x\in\mathcal{X}: l+\eta-2^{1-m}\le y(x)< l+\eta-2^{-m}\right\}, & \gamma=3.
\end{cases}
\]

The solutions in $\mathcal{S}_1$ have better mean performance, and their corresponding label in register $F$ is $1$. Conversely, the solutions in $\mathcal{S}_2$ have worse mean performance, and their corresponding label in register $F$ is $0$. The solutions in $\mathcal{S}_3$ lie in the intermediate region between the two thresholds. For these solutions, the labeling outcome is not deterministic: register $F$ may be either $1$ or $0$. This intermediate region is introduced to accommodate the estimation error of the quantum Monte Carlo mean estimation procedure.
Accordingly, the idealized output state of Algorithm~\ref{alg:qbso} is
\begin{equation*}
\begin{aligned}
|\psi(\mathcal A)\rangle
=
&\frac{1}{\sqrt{|\mathcal X|}}
\sum_{x\in \mathcal S_1}
|x\rangle_{\mathcal X} |\mathrm{data}_x\rangle_{\xi,Y}
|\psi_x\rangle_{A,P}|1\rangle_F \\
&+
\frac{1}{\sqrt{|\mathcal X|}}
\sum_{x\in \mathcal S_2}
|x\rangle_{\mathcal X} |\mathrm{data}_x\rangle_{\xi,Y}
|\psi_x\rangle_{A,P}|0\rangle_F \\
&+
\frac{1}{\sqrt{|\mathcal X|}}
\sum_{x\in \mathcal S_3}
|x\rangle_{\mathcal X} |\mathrm{data}_x\rangle_{\xi,Y}
\Bigl(
\lambda_{x,1}|\psi_{x,1}\rangle_{A,P}|1\rangle_F
+
\lambda_{x,0}|\psi_{x,0}\rangle_{A,P}|0\rangle_F
\Bigr)
\end{aligned}
\end{equation*}
for some $\lambda_{x,1},\lambda_{x,0}\in \mathbb{C}$ and states
$|\psi_x\rangle$, $|\psi_{x,1}\rangle$, and $|\psi_{x,0}\rangle$. Here
\[
|\mathrm{data}_x\rangle_{\xi,Y}
:=
\sum_{\xi_x\in\Xi}\sqrt{P(\xi_x)}\,|\xi_x\rangle_{\xi}|Y(x,\xi_x)\rangle_Y,
\]

Due to estimation error, the actual output state $|\phi(\mathcal A)\rangle$ may differ from the idealized state $|\psi(\mathcal A)\rangle$. Lemma~\ref{lemma:correctness_A} shows that $|\phi(\mathcal A)\rangle$ is within $\kappa/|\mathcal X|$ of $|\psi(\mathcal A)\rangle$, implying that the discrepancy between the two states can be controlled through the choice of the risk parameter $\kappa$. The lemma also provides an upper bound on the difference between the proportions of labeled solutions under these two states. The proof of Lemma~\ref{lemma:correctness_A} relies on the performance guarantee of the quantum Monte Carlo mean estimation procedure established in Lemma~\ref{lemma: qae}.

\begin{lemma}
\label{lemma:correctness_A}
Let $|\phi(\mathcal A)\rangle$ and $|\psi(\mathcal A)\rangle$ denote the actual and ideal output states of Algorithm~\ref{alg:qbso}, respectively. Then 
\[
\bigl\|\,|\phi(\mathcal A)\rangle-|\psi(\mathcal A)\rangle\,\bigr\|
\le \frac{\kappa}{|\mathcal{X}|}.
\] 
In particular, let
$
\Pi_F$
be the projector onto the flagged subspace $F=1$. Define
\[
p^{\mathrm{label}}:=\|\Pi_F |\phi(\mathcal A)\rangle\|^2,
\qquad
q^{\mathrm{label}}:=\|\Pi_F |\psi(\mathcal A)\rangle\|^2,
\]
to be the proportions of labeled solutions under the states $|\phi(\mathcal A)\rangle$ and $|\psi(\mathcal A)\rangle$, respectively. Then
\[
|p^{\mathrm{label}}-q^{\mathrm{label}}|
\le
\frac{2\kappa}{|\mathcal X|}.
\]
\end{lemma}

Lemma~\ref{lemma: complexity A} establishes the query complexity of Algorithm~\ref{alg:qbso}. The main oracle-querying operation in Algorithm~\ref{alg:qbso} is the quantum Monte Carlo mean estimation procedure. The result follows from the query complexity given in Lemma~\ref{lemma: qae} together with the boundedness of the random performance $Y(x,\xi_x)$.

\begin{lemma}[Query Complexity of Algorithm \ref{alg:qbso}]
\label{lemma: complexity A}
For a given precision $\eta\in(0,1)$ and risk level $\kappa\in(0,1)$, Algorithm~\ref{alg:qbso} uses
\[
O\!\left(\frac{1}{\eta}\log\frac{1}{\alpha}\right)
\]
queries to the global oracle $\mathcal Q$ and its inverse, where $\alpha = \frac{\kappa^2}{4|\mathcal{X}|^{3}}$.
\end{lemma}

Lemma~\ref{lemma: estimate-amplify} establishes the query complexity guarantees for the subprocedures $\textsc{Amplify}(\mathcal A,\delta)$ in Algorithm \ref{alg:grover} and $\textsc{Estimate}(\mathcal A,\epsilon,\delta)$ in Algorithm \ref{alg:confidence_interval}. The result is obtained by combining the query complexity bounds for amplitude amplification and amplitude estimation in Lemma~\ref{lem: amplitude amplify} and Lemma~\ref{lemma: ae} with the query complexity of Algorithm~\ref{alg:qbso} established in Lemma~\ref{lemma: complexity A}.

\begin{lemma}[Query Complexity of \textsc{Amplify} and \textsc{Estimate} Procedures]
\label{lemma: estimate-amplify}
Consider the procedure $\mathcal A=\mathcal A(\mathcal Q,\ell,\eta,\kappa)$ in Algorithm~\ref{alg:qbso}, and let $T_{\mathcal A}$ denote its query complexity. Assume that $\kappa=0.1\delta$. Suppose that there exists $x\in\mathcal{X}$ such that $y(x)\ge \ell+\eta-2^{-m}$. 
Then there exist quantum algorithms $\mathrm{Amplify}(\mathcal A,\delta)$ and
$\mathrm{Estimate}(\mathcal A,\varepsilon,\delta)$ with the following properties.
The algorithm $\mathrm{Amplify}(\mathcal A,\delta)$ uses 
\[
O\!\left(
\frac{T_{\mathcal A}}{\sqrt{p^{\mathrm{label}}}}
\log\frac{1}{\delta}
\right)
\]
queries to $\mathcal Q$ and its inverse, and outputs an element
$x\in \mathcal X$ satisfying
$
y(x)\ge \ell+\eta-2^{1-m}
$
with probability at least $1-\delta$.
The algorithm $\mathrm{Estimate}(\mathcal A,\varepsilon,\delta)$ uses 
\[
O\!\left(
\frac{T_{\mathcal A}}{\varepsilon\sqrt{p^{\mathrm{label}}}}
\log\frac{1}{\delta}
\right)
\]
queries to $\mathcal Q$ and its inverse, and outputs an estimate $r$ such that
$
|r-p^{\mathrm{label}}|\le \varepsilon p^{\mathrm{label}}
$
with probability at least $1-\delta$. Consequently, with probability at least $1-\delta$,
\[
r\in
\left(
(1-\varepsilon)\left(q^{\mathrm{label}}-\frac{2\kappa}{|\mathcal X|}\right),
\,
(1+\varepsilon)\left(q^{\mathrm{label}}+\frac{2\kappa}{|\mathcal X|}\right)
\right).
\]
\end{lemma}

We next analyze the performance of Algorithm~\ref{alg:confidence_interval}, which uses a binary-search-based procedure to construct an optimal region containing the optimal performance $y(x^\star)$ with high probability. Lemma~\ref{lem:shrink-single} shows that, at each iteration $t$, if the current region $\mathcal R_t$ contains $y(x^\star)$, then the updated region $\mathcal R_{t+1}$ produced by the binary-search procedure also contains $y(x^\star)$. It further gives the query complexity of each iteration. This correctness result relies on Lemma~\ref{lemma:correctness_A}. In particular, if the threshold parameter $r_t$ in Algorithm~\ref{alg:confidence_interval} is estimated accurately, then the procedure can identify the location of $y(x^\star)$ and eliminate suboptimal regions accordingly.
 
\begin{lemma}[Per-Iteration Correctness and Query Complexity of Algorithm \ref{alg:confidence_interval}]
\label{lem:shrink-single}
For each iteration \(t\), let \(\mathcal R_{t+1}\) be the interval returned from the current interval \(\mathcal R_t\). Then, with probability at least \(1-\delta_t\),
\[
y(x^\star)\in \mathcal R_t
\;\Longrightarrow\;
y(x^\star)\in \mathcal R_{t+1}.
\]
Moreover,
$
|\mathcal R_{t+1}|
\le
\frac{11}{16}\,|\mathcal R_t|.
$
Finally, iteration \(t\) uses
\[
O\!\left(
\frac{\sqrt{|\mathcal X|+1}}{\eta_t}\,
\operatorname{polylog}\!\left(\frac{|\mathcal X|+1}{\delta_t}\right)
\right)
\]
queries to \(\mathcal Q\) and its inverse.
\end{lemma}

Lemma~\ref{lem:confidence-interval} establishes the overall performance of Algorithm~\ref{alg:confidence_interval}. Since $y(x^\star) \in \mathcal{R}_0$, repeated application of Lemma~\ref{lem:shrink-single} implies that $y(x^\star)$ remains in the region throughout the procedure, and therefore belongs to the final returned region $\mathcal{R}$. The overall failure probability is controlled to be at most $\delta/2$.
Moreover, we show that the width of the region decreases geometrically:
\begin{equation*}
    |\mathcal{R}_t| \le \left(\frac{11} {16}\right)^t |\mathcal{R}_0|.
\end{equation*}
This implies that the number of iterations is bounded logarithmically, and in particular, the width of the returned region $\mathcal{R}$ is less than $\epsilon/2$. The lemma also provides an upper bound on the total query complexity of the algorithm.

\begin{lemma}[Correctness and Query Complexity of Algorithm \ref{alg:confidence_interval}]
\label{lem:confidence-interval}
Let \(\mathcal{R}\) denote the interval returned by Algorithm~\ref{alg:confidence_interval}. Then, with probability at least \(1-\frac{\delta}{2}\), 
\[
y(x^\star) \in \mathcal{R},\quad
|\mathcal{R}|<\frac{\epsilon}{2}.
\]
Moreover, Algorithm~\ref{alg:confidence_interval} terminates after at most
\[
T=
\left\lceil
\log_{16/11}\frac{2}{\epsilon}
\right\rceil
\]
iterations, and its total query complexity is
\[
O\!\left(
\frac{\sqrt{|\mathcal{X}|}}{\epsilon}\,
\operatorname{polylog}\!\left(\frac{|\mathcal{X}|}{\delta\epsilon}\right)
\right).
\]
\end{lemma}

Based on these results, we are now ready to establish the performance guarantee of Algorithm~\ref{alg:grover}. Theorem~\ref{thm:gas} shows that the SOGAS algorithm returns an $\epsilon$-optimal solution with probability at least $1-\delta$ using $\tilde{O}(\sqrt{|\mathcal{X}|})$ queries. This proves that the algorithm satisfies an $(\epsilon,\delta)$-PAC guarantee and demonstrates a quadratic quantum speedup, since the worst-case query complexity on a classical computer is $\tilde{O}(|\mathcal{X}|)$.
\begin{theorem}[Correctness and Query Complexity of Algorithm \ref{alg:grover}]
\label{thm:gas}
For any constant optimality gap $\epsilon \in (0,1)$ and risk level $\delta \in (0,1)$, Algorithm~\ref{alg:grover} outputs an $\epsilon$-optimal arm $\hat{x}^\star$ with probability at least $1-\delta$, that is,
$
y(\hat{x}^\star) \ge y(x^\star)-\epsilon.
$
Moreover, its query complexity is $\tilde{O}(\sqrt{|\mathcal{X}|})$, where $\tilde{O}(\cdot)$ suppresses polylogarithmic factors.
\end{theorem}

The analysis is based on the event that the optimal region $\mathcal{R}$ contains the optimal performance $y(x^\star)$ and has width less than $\epsilon/2$. On this event, under the ideal labeling rule of Algorithm \ref{alg:qbso}, all solutions whose mean performances lie in $\mathcal{R}$ are labeled as target solutions. In addition, we introduce a buffer region, illustrated in Figure \ref{alg:grover}, to accommodate the estimation error. Some solutions in the buffer region may also be labeled. Since the width of the buffer region is also guaranteed to be less than $\epsilon/2$, every solution labeled under the ideal rule is $\epsilon$-optimal. Finally, amplitude amplification is applied to return one of the labeled solutions with high probability.

\section{Numerical Experiments}
\label{sec: experiment}
In this section, we conduct numerical experiments to evaluate the empirical performance of the proposed SOGAS algorithm for discrete simulation optimization. We examine its performance under different problem scales, optimality gaps, and performance distributions.

To enable a fair comparison when evaluating quantum advantage, we introduce a classical counterpart, referred to as classical SOGAS (CSOGAS). CSOGAS follows the same overall procedure as the quantum SOGAS algorithm, but replaces the quantum amplitude estimation subroutine with a classical sample-average estimator. In the classical setting, generating one random sample is counted as one query to a classical oracle. To achieve a comparable confidence guarantee, we combine Hoeffding’s inequality with a heuristic rule for determining the required query budget. Specifically, we first impose a small minimum sampling budget based on a fixed fraction of the Hoeffding bound, and then increase the number of oracle queries adaptively in batches according to the empirical variance until the stopping criterion is satisfied. This strategy is less conservative than using Hoeffding’s inequality alone to determine the query budget. As a result, any observed performance difference can be attributed primarily to the estimation procedure, rather than to differences in the overall algorithmic framework.

We implement the SOGAS algorithm using Qiskit-based simulation \citep{javadi2024quantum}. In our experiments, the exact quantum flagging procedure in Algorithm~\ref{alg:qbso} is not directly implementable, because classically simulating its fully coherent evolution is computationally prohibitive. To make the study feasible, we therefore adopt an approximate implementation. Specifically, for each solution, we use quantum mean estimation to estimate its mean performance, then measure the output, and assign the corresponding flag classically. This yields a hybrid quantum-classical workflow that can be simulated on a classical computer.

This approximation has an important limitation. The intermediate measurement collapses the quantum state and destroys the coherent superposition required by the exact algorithm, which weakens the ideal quadratic speedup. Nevertheless, the approximate quantum algorithm still significantly outperforms the classical benchmark in our experiments. Since the exact fully coherent quantum algorithm is theoretically stronger than this approximate implementation, these results still provide evidence of a genuine quantum advantage.

Throughout this section, the query complexity and probability of correct selection are estimated from 30 independent replications, and the corresponding 95\% confidence intervals are reported. The overall risk level is set to $\delta=0.05$. Because the algorithm is designed conservatively to ensure statistical validity, both SOGAS and CSOGAS attain 100\% correct selection in all experiments. We therefore focus on comparing their empirical query complexities.

We first evaluate the algorithm under different problem sizes. In this experiment, the number of candidate solutions varies over $|\mathcal{X}|\in\left\{5,10,15,20,25\right\}$, while the target optimality gap is fixed at $\epsilon=0.1$. For each value of $|\mathcal{X}|$, solution performance is modeled by a Bernoulli distribution, with the unknown mean of each solution generated within a narrow range of approximately $0.78$ to $0.85$. Consequently, the candidate solutions have very similar performance levels, making the simulation optimization problem nontrivial.

\begin{figure}
    \centering
    \includegraphics[width=0.6\linewidth]{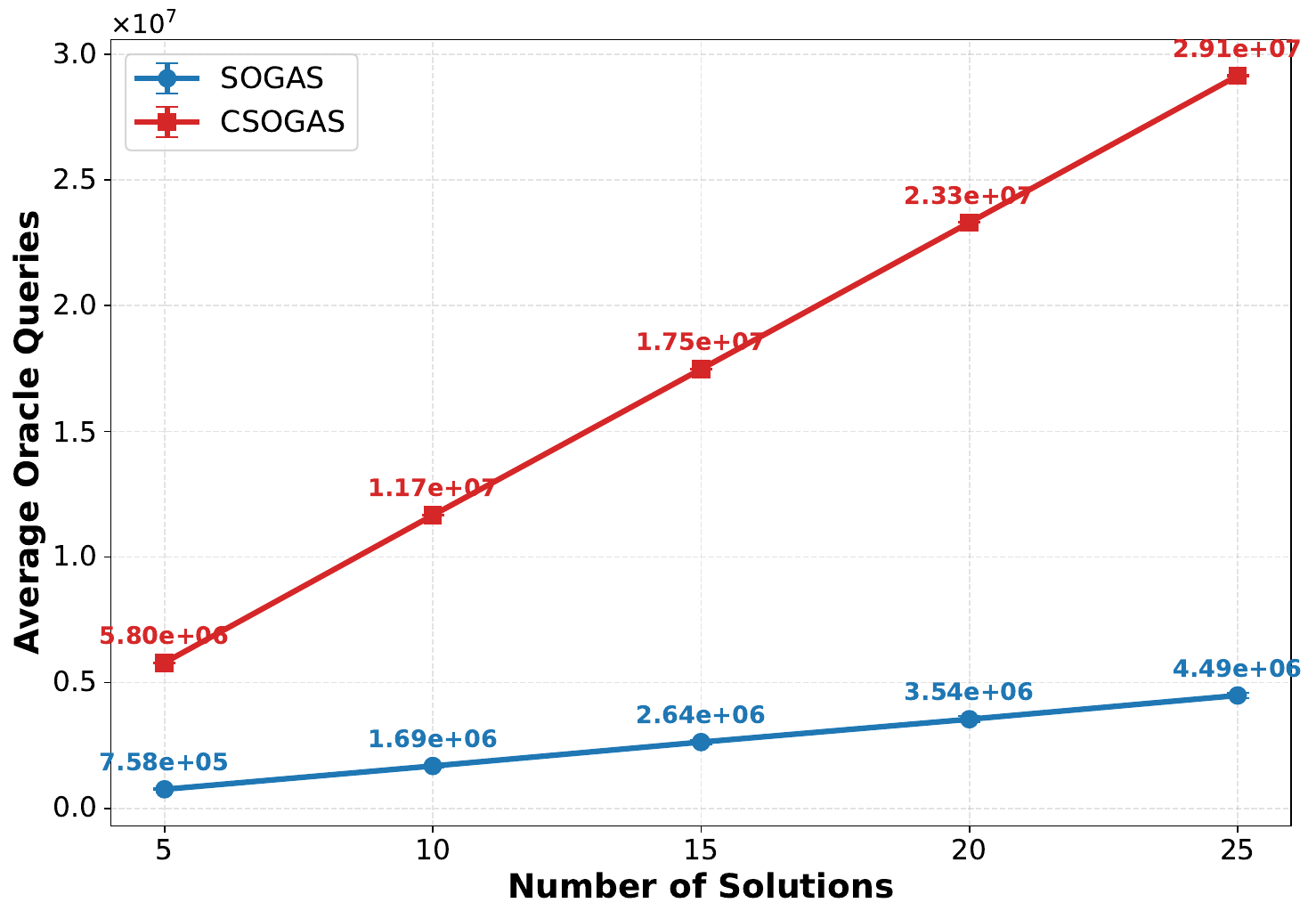}
    \caption{Empirical Query Complexity for Different Problem Sizes}
    \label{fig:problem size}
\end{figure}

Figure \ref{fig:problem size} shows that the empirical query complexity of both SOGAS and CSOGAS increases as the number of solutions grows, but the increase for SOGAS is much more gradual. Across all tested problem sizes, SOGAS uses only a small fraction of the queries required by CSOGAS, achieving an approximately $6$--$8\times$ reduction in query complexity. This suggests that the quantum approach scales more favorably with problem size and remains substantially more query-efficient than the classical benchmark, providing clear evidence of quantum advantage.

It is worth noting that these results are obtained under an approximate implementation of Algorithm \ref{alg:qbso}. In particular, the intermediate measurement collapses the quantum state and destroys the coherence needed to realize the ideal quadratic speedup. As a result, the empirical query complexity exhibits an approximately linear dependence on the problem size, which is consistent with what one would expect from this hybrid quantum-classical implementation. Even with this limitation, however, SOGAS still significantly outperforms the classical benchmark. Since the exact, fully coherent quantum algorithm is theoretically stronger than the approximate version used in our experiments, its actual query complexity is expected to be even lower. Therefore, these results already provide strong empirical evidence of a genuine quantum advantage.

We next evaluate the algorithm under different values of the optimality gap $\epsilon$. In this experiment, the inverse gap is varied over $1/\epsilon \in \left\{5,10,15,20,25\right\}$, while the number of candidate solutions is fixed at $|\mathcal{X}|=25$. The underlying Bernoulli instance is the same as that used in the $|\mathcal{X}|=25$ setting of the problem-size experiment, with only the optimality gap adjusted across different cases.

\begin{figure}
\centering
\includegraphics[width=0.6\linewidth]{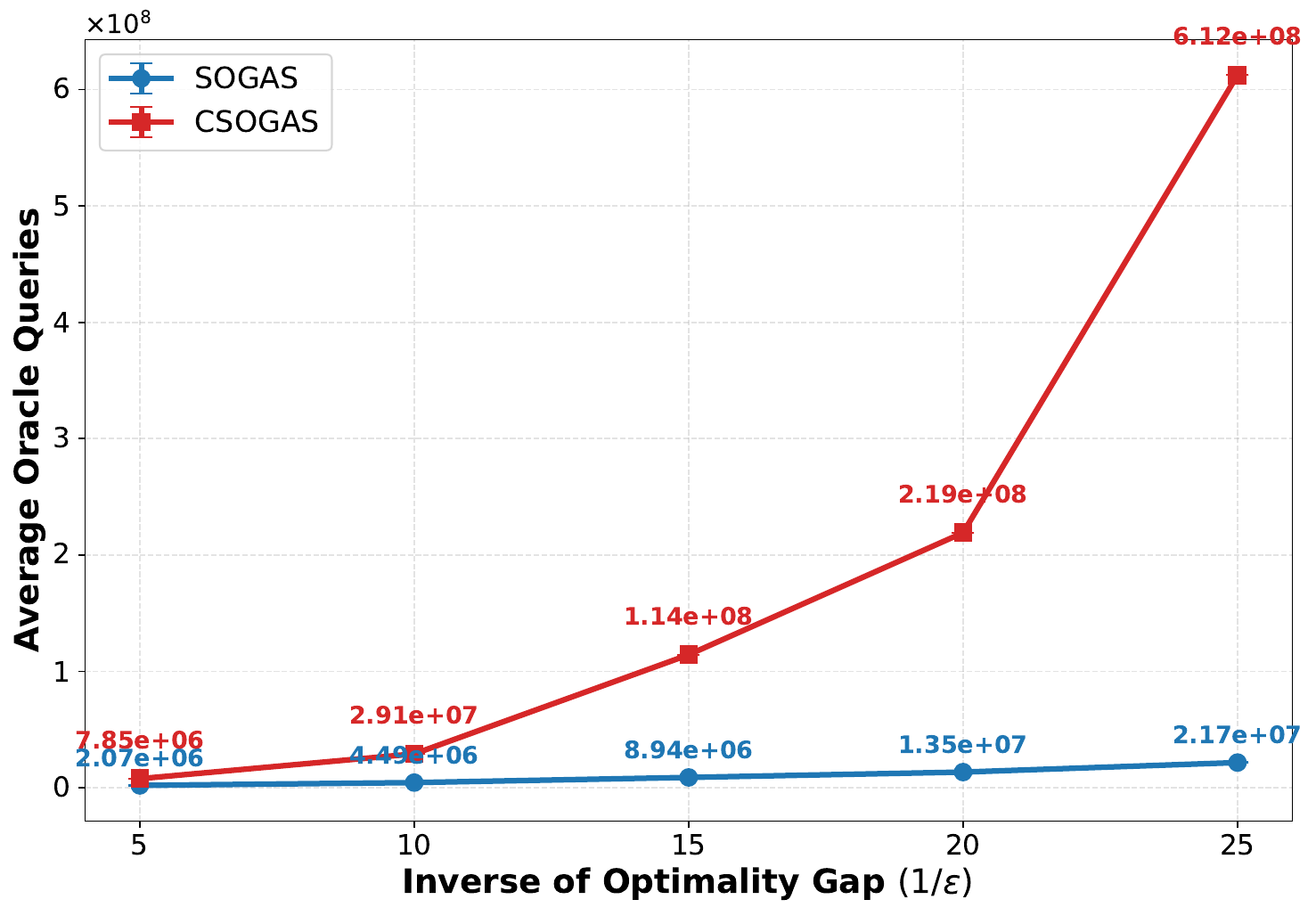}
\caption{Empirical Query Complexity for Different Optimality Gap}
\label{fig:optimality gap}
\end{figure}

Figure \ref{fig:optimality gap} shows that the empirical query complexity of both SOGAS and CSOGAS increases as $1/\epsilon$ becomes larger, but the increase for SOGAS is noticeably slower. For all tested values of $\epsilon$, SOGAS consistently uses far fewer queries than CSOGAS, indicating a substantial improvement in query efficiency. In particular, as $1/\epsilon$ increases, the query complexity of CSOGAS exhibits an approximately quadratic dependence, whereas that of SOGAS grows nearly linearly. This contrast in scaling suggests a clear quadratic speedup of the quantum algorithm over the classical benchmark. Overall, these results show that SOGAS remains significantly more query-efficient across different optimality-gap settings and provide strong empirical support for the practical superiority of the quantum algorithm.

Finally, we evaluate the algorithm under different performance distributions. In this experiment, the number of candidate solutions is fixed at $|\mathcal{X}|=10$, and the target optimality gap is fixed at $\epsilon=0.1$. We consider three distribution families: Gaussian, Uniform, and Exponential. For each family, the mean performance of each solution is generated within the interval $[0,1]$. More specifically, in the Gaussian case, the underlying centers are evenly spaced, producing means approximately from $0.46$ to $0.80$; in the Uniform case, the resulting means range from about $0.50$ to $0.81$; and in the Exponential case, they range from about $0.62$ to $0.83$. In all three settings, the solution mean performances are chosen to be close to one another, so that the resulting simulation optimization problems remain nontrivial.

On the quantum side, each continuous distribution is approximated by a discrete distribution with finite support on $[0,1]$. Specifically, we use $3$ uncertainty qubits, corresponding to $8$ discretization points, to encode each distribution into a quantum state. The probability masses on these eight support points are loaded into the amplitudes, and the corresponding performance values are then mapped to the amplitude of the objective qubit for amplitude estimation. In this way, the quantum algorithm represents the original continuous distribution through a discretized oracle approximation. 

\begin{figure}
\centering
\includegraphics[width=1.0\linewidth]{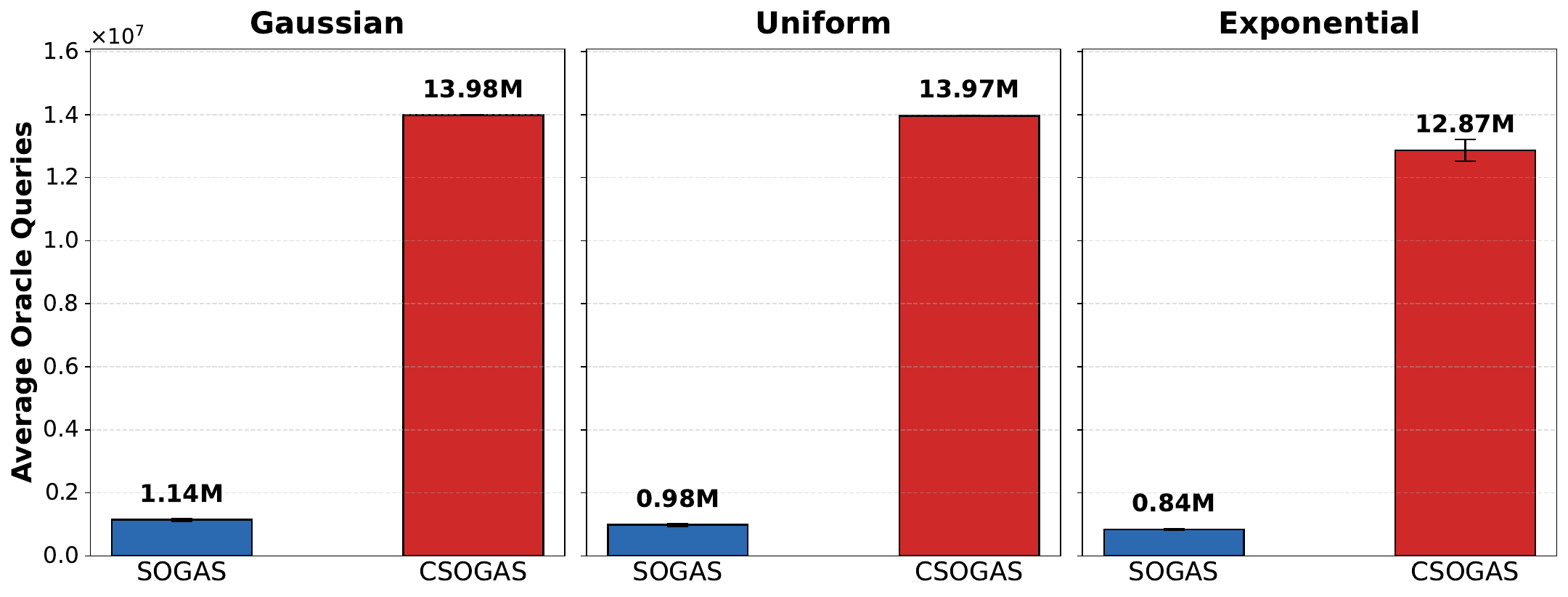}
\caption{Empirical Query Complexity for Different Distributions}
\label{fig:distributions}
\end{figure}

Figure \ref{fig:distributions} compares the empirical query complexity of SOGAS and CSOGAS under three different performance distributions. In all three cases, SOGAS requires substantially fewer oracle queries than CSOGAS. Across these distributions, SOGAS achieves an approximately $12$-$15\times$ speedup over CSOGAS. These results show that the quantum algorithm consistently outperforms the classical benchmark across different distribution families while maintaining much lower query complexity. This suggests that the advantage of SOGAS is robust to the underlying performance distribution and provides strong empirical support for the superior query efficiency of the quantum approach.

\section{Conclusion}
\label{conclusion}
This paper investigates the use of quantum computing for discrete simulation optimization. We propose a Grover-search-based algorithm, SOGAS, and show that it returns a near-optimal solution with high probability while achieving a quadratic speedup in query complexity. Numerical results further support the quantum advantage. Promising directions for future research include extending the proposed approach to continuous simulation optimization and to more complex settings, such as nested simulation and correlated sampling noise. 

Overall, this work opens the door to applying quantum computing to accelerate simulation optimization. It shows that algorithmic techniques in the simulation literature can benefit from the computational paradigm of quantum computers, leading to new problem formulations, algorithmic frameworks, and theoretical analyses.


%
\begin{APPENDIX}{Electronic Companion for “Quantum Grover Adaptive Search for Discrete Simulation Optimization”}
\section{Proof of Lemma \ref{lem: gapped estimation}.} 
Fix any $x\in\mathcal X$. Let $\hat{\mu}_x$ denote the estimate produced by
$\textsc{QAE}(2^{-m}/3,\alpha)$.
By Lemma~\ref{lemma: qae},
\[
\mathbb{P}\left(\left|\hat{\mu}_x-\mathbb{E}[Y(x,\xi_x)]\right|\le \frac{2^{-m}}{3}\right)\ge 1-\alpha.
\]
Since the whole procedure is unitary, for each fixed $x$ its output on registers $A$ and $P$ can be written as
\[
\lambda_{x,0}\ket{0}_A\ket{\gamma_{x,0}}_{P}
+
\lambda_{x,1}\ket{1}_A\ket{\gamma_{x,1}}_{P},
\]
for some normalized states $\ket{\gamma_{x,0}}_{P}$ and $\ket{\gamma_{x,1}}_{P}$, where
\[
\lambda_{x,0},\lambda_{x,1}\in[0,1],
\qquad
\lambda_{x,0}^2+\lambda_{x,1}^2=1.
\]
Here we take $\lambda_{x,0}$ and $\lambda_{x,1}$ to be nonnegative real numbers by absorbing any phase factors into $\ket{\gamma_{x,0}}_{P}$ and $\ket{\gamma_{x,1}}_{P}$. The register $F$ is unchanged throughout this procedure, and hence remains in the state $\ket{0}_{F}$.

Now suppose
$
\mathbb{E}[Y(x,\xi_x)]\ge \ell+\eta-2^{-m}.
$ On the event $|\hat{\mu}_x-\mathbb{E}[Y(x,\xi_x)]|\le 2^{-m}/3$, we have 
\[
\hat{\mu}_x
\ge
\mathbb{E}[Y(x,\xi_x)]-\frac{1}{3}2^{-m}
\ge
\ell+\eta-\frac{4}{3}2^{-m}.
\]
Since
\[
\ell+\eta-\frac{4}{3}2^{-m}
>
\ell+\eta-\frac{3}{2}2^{-m},
\]
the estimate is closer to $\ell+\eta-2^{-m}$ than to $\ell+\eta-2^{1-m}$, so the coherent comparison writes label $\ket{1}_A$. Therefore, the wrong label $\ket{0}_A$ can occur with probability at most $\alpha$. Hence
$
\lambda_{x,0}^2\le \alpha.
$ Similarly, suppose $
\mathbb{E}[Y(x,\xi_x)]< \ell+\eta-2^{1-m}.
$ On the event $|\hat{\mu}_x-\mathbb{E}[Y(x,\xi_x)]|\le 2^{-m}/3$, we have \[
\hat{\mu}_x
\le
\mathbb{E}[Y(x,\xi_x)]+\frac{1}{3}2^{-m}
<
\ell+\eta-\frac{5}{3}2^{-m}.
\]Since
\[
\ell+\eta-\frac{5}{3}2^{-m}
<
\ell+\eta-\frac{3}{2}2^{-m},
\]the estimate is closer to $\ell+\eta-2^{1-m}$ than to $\ell+\eta-2^{-m}$, so the coherent comparison writes label $\ket{0}_A$. Therefore the wrong label $\ket{1}_A$ can occur with probability at most $\alpha$. Hence
$
\lambda_{x,1}^2\le \alpha.
$ Applying this argument coherently for all $x\in\mathcal X$ yields the claimed global form of the output state.

\section{Proof of Lemma \ref{lemma:correctness_A}}
This proof follows the argument used in the proof of Lemma 6 in \citep{wang2021quantum}. The initial state of Algorithm~\ref{alg:qbso} is
\[
|\Phi_0\rangle
=
\frac{1}{\sqrt{|\mathcal X|}}
\sum_{x\in\mathcal X}
|x\rangle_{\mathcal{X}}
|\mathrm{data}_x\rangle_{\xi,Y}
|0\rangle_A|0\rangle_P|0\rangle_F.
\]

Since $\textsc{QAE}(2^{-m}/3,\alpha)$ is applied coherently conditioned on register $\mathcal X$, it suffices to analyze the state for each fixed $x\in\mathcal X$. We distinguish three cases.

\medskip
\noindent
\textbf{Case 1: $x\in \mathcal S_3$.}
In this case, Lemma~\ref{lem: gapped estimation} does not impose any additional constraint on the label probabilities. Hence, after step $7$ of Algorithm \ref{alg:qbso}, the state can simply be written as a superposition of the two flag branches,
\[
\lambda_{x,1}|\psi_{x,1}\rangle_{A,P}|1\rangle_F
+
\lambda_{x,0}|\psi_{x,0}\rangle_{A,P}|0\rangle_F,
\]
for some states $|\psi_{x,1}\rangle$ and $|\psi_{x,0}\rangle$.

\medskip
\noindent
\textbf{Case 2: $x\in \mathcal S_1$.}
Then $\lambda_{x,0}^2\le \alpha$, so after the controlled-NOT gate the state has the form
\[
\lambda_{x,1}|\psi_{x,1}\rangle_{A,P}|1\rangle_F
+
\lambda_{x,0}|\psi_{x,0}\rangle_{A,P}|0\rangle_F,
\]
with $\lambda_{x,1}=\sqrt{1-\lambda_{x,0}^2}$. Hence
\[
\Bigl\|
\lambda_{x,1}|\psi_{x,1}\rangle_{A,P}|1\rangle_F
+
\lambda_{x,0}|\psi_{x,0}\rangle_{A,P}|0\rangle_F
-
|\psi_{x,1}\rangle_{A,P}|1\rangle_F
\Bigr\|
\le
|1-\lambda_{x,1}|+\lambda_{x,0}.
\]
Since $\lambda_{x,0}\le \sqrt{\alpha}$ and
\[
1-\lambda_{x,1}
=
1-\sqrt{1-\lambda_{x,0}^2}
\le \lambda_{x,0}^2
\le \alpha
\le \sqrt{\alpha},
\]
the above is at most
$
2\sqrt{\alpha}.
$
Therefore, the resulting state is $2\sqrt{\alpha}$-close to a state of the form
$
|\psi_x\rangle_{A,P}|1\rangle_F.
$

\medskip
\noindent
\textbf{Case 3: $x\in \mathcal S_2$.}
Similarly, $\lambda_{x,1}^2\le \alpha$, so after the controlled-NOT gate the resulting state is $2\sqrt{\alpha}$-close to a state of the form
$
|\psi_x\rangle_{A,P}|0\rangle_F.
$

\medskip
Combining the above three cases, for each fixed $x$ the final state produced from the basis component
\[
|x\rangle_X |\mathrm{data}_x\rangle_{\xi,Y}|0\rangle_A|0\rangle_P|0\rangle_F
\]
is $2\sqrt{\alpha}$-close to the corresponding component appearing in $|\psi(\mathcal A)\rangle$. Therefore, by the triangle inequality,
\[
\bigl\|\,|\phi(\mathcal A)\rangle-|\psi(\mathcal A)\rangle\,\bigr\|
\le
\frac{1}{\sqrt{|\mathcal X|}}
\sum_{x\in\mathcal X} 2\sqrt{\alpha}
=
2\sqrt{\alpha|\mathcal X|}.
\]
Choosing
$
\alpha=\frac{\kappa^2}{4|\mathcal X|^3}
$
gives
$
2\sqrt{\alpha|\mathcal X|}
=
\frac{\kappa}{|\mathcal X|}.
$
Hence $|\phi(\mathcal A)\rangle$ is within $\kappa/|\mathcal X|$ of $|\psi(\mathcal A)\rangle$.

Finally,
\[
|p^{\mathrm{label}}-q^{\mathrm{label}}|
=
\bigl(\sqrt{p^{\mathrm{label}}}+\sqrt{q^{\mathrm{label}}}\bigr)
\cdot
\bigl|\sqrt{p^{\mathrm{label}}}-\sqrt{q^{\mathrm{label}}}\bigr|
\le
2\|\Pi_F(|\phi(\mathcal A)\rangle-|\psi(\mathcal A)\rangle)\|
\]
\[
\le
2\bigl\|\,|\phi(\mathcal A)\rangle-|\psi(\mathcal A)\rangle\,\bigr\|
\le
\frac{2\kappa}{|\mathcal X|}.
\]
This completes the proof.

\section{Proof of Lemma \ref{lemma: complexity A}}

Algorithm~\ref{alg:qbso} performs exactly one coherent call to
$
\textsc{Qae}(2^{-m}/3,\alpha)
$
conditioned on the register $\mathcal X$, followed by a coherent comparison and a controlled-NOT gate. The comparison step and the controlled-NOT gate do not query $\mathcal Q$ or $\mathcal Q^\dagger$.

For each fixed $x\in\mathcal X$, the oracle $\mathcal Q_x$ prepares the random variable $Y(x,\xi_x)$. By Lemma~\ref{lemma: qae}, the procedure
$
\textsc{Qae}(\epsilon,\delta)
$
uses
\[
O\!\left(\frac{\sigma}{\epsilon}\log\frac{1}{\delta}\right)
\]
queries to the underlying oracle and its inverse, where $\sigma$ is an upper bound on the standard deviation of $Y(x,\xi_x)$.

Substituting
$
\epsilon=\frac{2^{-m}}{3},
\delta=\alpha,
$
we obtain that Algorithm~\ref{alg:qbso} uses
$
O(\sigma\,2^m\log\frac{1}{\alpha})
$
queries to $\mathcal Q$ and $\mathcal Q^\dagger$.

Finally, since
$
m=\left\lceil \log_2\frac{1}{\Delta}\right\rceil+2,
$
we have
$
2^m=\Theta(\frac{1}{\Delta}).
$
Hence, the query complexity is
\[
O\!\left(\frac{\sigma}{\Delta}\log\frac{1}{\alpha}\right)
=
O\!\left(\frac{\sigma}{\eta}\log\frac{1}{\alpha}\right),
\]
because $\Delta=2\eta$.
Since $Y(x,\xi_x)\in[0,1]$, its standard deviation is at most $1/2$, and therefore
\[
O\!\left(\frac{\sigma}{\eta}\log\frac{1}{\alpha}\right)
=
O\!\left(\frac{1}{\eta}\log\frac{1}{\alpha}\right).
\]

\section{Proof of Lemma \ref{lemma: estimate-amplify}} 
By Lemma~\ref{lemma: complexity A}, we have
\[
T_{\mathcal A}
=
O\!\left(\frac{1}{\eta}\log\frac{1}{\alpha}\right).
\]
Consider first $\mathrm{Amplify}(\mathcal A,\delta)$. By amplitude amplification \citep{brassard2000quantum} in Lemma \ref{lem: amplitude amplify}, a quantum procedure with query complexity
$T_{\mathcal A}$ and probability mass $p^{\mathrm{label}}$ on the labeled subspace can be amplified so that a flagged state is produced with probability at least $1-\delta/2$ using
\[
O\!\left(
\frac{T_{\mathcal A}}{\sqrt{p^{\mathrm{label}}}}
\log\frac{1}{\delta}
\right)
\]
queries.

Let
\[
\mathcal G:=\{x\in\mathcal X:\ y(x)\ge l+\eta-2^{1-m}\},
\qquad
\mathcal B:=\mathcal X\setminus\mathcal G,
\]
and define
\[
\Pi_{\mathrm{bad}}
:=
\Bigl(\sum_{x\in\mathcal B}|x\rangle\!\langle x|_{\mathcal X}\Bigr)\otimes I_{\xi,Y,A,P}\otimes |1\rangle\!\langle 1|_F,
\]
which is the projector onto the subspace spanned by states with $x\in\mathcal{B}$ and $F=1$.
By construction of the idealized state $|\psi(\mathcal A)\rangle$, its labeled subspace is supported only on $\mathcal G$. Hence
\[
\Pi_{\mathrm{bad}}|\psi(\mathcal A)\rangle=0.
\]
Therefore, by Lemma~\ref{lemma:correctness_A},
\[
\|\Pi_{\mathrm{bad}}|\phi(\mathcal A)\rangle\|
=
\|\Pi_{\mathrm{bad}}(|\phi(\mathcal A)\rangle-|\psi(\mathcal A)\rangle)\|
\le
\bigl\|\,|\phi(\mathcal A)\rangle-|\psi(\mathcal A)\rangle\,\bigr\|
\le
\frac{\kappa}{|\mathcal X|},
\]
which implies
\[
\|\Pi_{\mathrm{bad}}|\phi(\mathcal A)\rangle\|^2
\le
\frac{\kappa^2}{|\mathcal X|^2}.
\]
Conditioned on successful amplification, the output distribution is the same as measuring the normalized labeled component
\[
\frac{\Pi_F|\phi(\mathcal A)\rangle}{\|\Pi_F|\phi(\mathcal A)\rangle\|}.
\]
Hence the conditional probability of outputting a bad solution is at most
\[
\frac{\|\Pi_{\mathrm{bad}}|\phi(\mathcal A)\rangle\|^2}{p^{\mathrm{label}}}
\le
\frac{\kappa^2}{|\mathcal X|^2\,p^{\mathrm{label}}}.
\]
Define
\[
\mathcal S_1:=\{x\in\mathcal X:\ y(x)\ge l+\eta-2^{-m}\}.
\]
By construction of the idealized state, every \(x\in\mathcal S_1\) belongs to the labeled subspace under the ideal labeling rule. Therefore,
\[
q^{\mathrm{label}}\ge \frac{|\mathcal S_1|}{|\mathcal X|}.
\]
Using Lemma~\ref{lemma:correctness_A}, we obtain
\[
p^{\mathrm{label}}
\ge
q^{\mathrm{label}}-\frac{2\kappa}{|\mathcal X|}
\ge
\frac{|\mathcal S_1|-2\kappa}{|\mathcal X|}.
\]
In particular, if \(|\mathcal S_1|\ge 1\), then
\[
p^{\mathrm{label}}\ge \frac{1-2\kappa}{|\mathcal X|}.
\]

Therefore,
\[
\frac{\kappa^2}{|\mathcal X|^2\,p^{\mathrm{label}}}
\le
\frac{\kappa^2}{|\mathcal X|(1-2\kappa)}.
\]

Thus, when \(|\mathcal S_1|\ge 1\), the overall probability that $\mathrm{Amplify}(\mathcal A,\delta)$ outputs an element
$x\in\mathcal X$ satisfying
$
y(x)\ge l+\eta-2^{1-m}
$
is at least
\[
\left(1-\frac{\delta}{2}\right)\left(1-\frac{\kappa^2}{|\mathcal X|(1-2\kappa)}\right).
\]

Now set
$
\kappa=0.1\,\delta.
$ Since $|\mathcal X|\ge 1$ and \(\delta\in(0,1)\), we have \(1-2\kappa\ge 0.8\), and hence
\[
\frac{\kappa^2}{|\mathcal X|(1-2\kappa)}
\le
\frac{\kappa^2}{0.8}
=
\frac{0.01\,\delta^2}{0.8}
\le
0.0125\,\delta
<
\frac{\delta}{2}.
\]
Hence
\[
\left(1-\frac{\delta}{2}\right)\left(1-\frac{\kappa^2}{|\mathcal X|(1-2\kappa)}\right)
\ge
1-\frac{\delta}{2}-\frac{\kappa^2}{|\mathcal X|(1-2\kappa)}
\ge
1-\delta.
\]

Therefore, whenever \(|\mathcal S_1|\ge 1\), $\mathrm{Amplify}(\mathcal A,\delta)$ outputs an element
$x\in\mathcal X$ satisfying
$
y(x)\ge l+\eta-2^{1-m}
$
with probability at least $1-\delta$.

Next consider $\mathrm{Estimate}(\mathcal A,\varepsilon,\delta)$.
By amplitude estimation in Lemma \ref{lemma: ae}, one can estimate the labeled solution proportion
$p^{\mathrm{label}}$ to multiplicative accuracy $\varepsilon$ with probability at least
$1-\delta$ using
\[
O\!\left(
\frac{T_{\mathcal A}}{\varepsilon\sqrt{p^{\mathrm{label}}}}
\log\frac{1}{\delta}
\right)
\]
queries. Hence the output $r$ satisfies
$
|r-p^{\mathrm{label}}|\le \varepsilon p^{\mathrm{label}}
$ with probability at least $1-\delta$. 
Finally, by Lemma~\ref{lemma:correctness_A},
\[
\left|p^{\mathrm{label}}-q^{\mathrm{label}}\right|
\le
\frac{2\kappa}{|\mathcal X|}.
\]
Therefore,
\[
q^{\mathrm{label}}-\frac{2\kappa}{|\mathcal X|}
\le
p^{\mathrm{label}}
\le
q^{\mathrm{label}}+\frac{2\kappa}{|\mathcal X|}.
\]
Combining this with
\[
(1-\varepsilon)p^{\mathrm{label}}<r<(1+\varepsilon)p^{\mathrm{label}},
\]
we obtain
\[
(1-\varepsilon)\left(q^{\mathrm{label}}-\frac{2\kappa}{|\mathcal X|}\right)
<
r
<
(1+\varepsilon)\left(q^{\mathrm{label}}+\frac{2\kappa}{|\mathcal X|}\right).
\]
This completes the proof.

\section{Proof of Lemma \ref{lem:shrink-single}}
 For a fixed iteration \(t\), define $\mathcal{S}^{\mathrm{label}}_t$ as the set of labeled solutions at iteration $t$. Let
\[
\bar{\mathcal X}:=\mathcal X\cup\{x_a\},
\qquad
|\bar{\mathcal X}|=|\mathcal X|+1,
\]
where the auxiliary solution \(x_a\) satisfies \(y(x_a)=1\).
Since
$
\ell_t+\eta_t-2^{-m_t}\le 1,
$
the auxiliary solution is always labeled in the augmented instance, that is,
$
x_a\in \mathcal S^{\mathrm{label}}_t.
$

Let \(p_t^{\mathrm{label}}\) and \(q_t^{\mathrm{label}}\) denote the actual and ideal labeled probabilities of \(\mathcal A_t\), respectively. Since $\kappa_t = 0.1\delta_t$, by Lemma~\ref{lemma:correctness_A},
\begin{equation}
\label{eq:prob-diff-t}
\bigl|p_t^{\mathrm{label}}-q_t^{\mathrm{label}}\bigr|
\le
\frac{0.2\,\delta_t}{|\bar{\mathcal X}|}.
\end{equation}

Hence
\[
\bigl|p_t^{\mathrm{label}}-q_t^{\mathrm{label}}\bigr|
\le
\frac{0.2\delta_t}{|\bar{\mathcal X}|}
<
\frac{0.2}{|\bar{\mathcal X}|}.
\]

Applying Lemma~\ref{lemma: estimate-amplify} to
$
r_t=\textsc{Estimate}(\mathcal A_t,0.1,\delta_t),
$
we obtain that, with probability at least \(1-\delta_t\),
\begin{equation}
\label{eq:conf-bound-t}
0.9\left(q_t^{\mathrm{label}}-\frac{0.2}{|\bar{\mathcal X}|}\right)
<
r_t
<
1.1\left(q_t^{\mathrm{label}}+\frac{0.2}{|\bar{\mathcal X}|}\right).
\end{equation}
We condition on the event that \eqref{eq:conf-bound-t} holds and show that the returned interval \(\mathcal R_{t+1}\) contains \(y(x^\star)\).

\medskip
\noindent
\textbf{Case 1:}
$
y(x^\star)<\ell_t+\eta_t-2^{1-m_t}.
$

Then no original solution in \(\mathcal X\) is labeled, so the only labeled solution in the augmented instance is \(x_a\). Hence
$
q_t^{\mathrm{label}}=\frac{1}{|\bar{\mathcal X}|}.
$
By \eqref{eq:conf-bound-t},
\[
r_t
<
1.1\left(\frac{1}{|\bar{\mathcal X}|}+\frac{0.2}{|\bar{\mathcal X}|}\right)
=
\frac{1.32}{|\bar{\mathcal X}|}
<
\frac{3}{2|\bar{\mathcal X}|}.
\]
Therefore the algorithm takes the second branch and returns
$
\mathcal R_{t+1}
=
[a_t,\;\ell_t+\eta_t-2^{-m_t}].
$
Since
\[
y(x^\star)<\ell_t+\eta_t-2^{1-m_t}<\ell_t+\eta_t-2^{-m_t},
\]
we conclude that \(y(x^\star)\in \mathcal R_{t+1}\).

\medskip
\noindent
\textbf{Case 2:}
$
y(x^\star)\ge \ell_t+\eta_t-2^{-m_t}.
$

Then \(x^\star\) is labeled, and together with \(x_a\) there are at least two labeled solutions in the augmented instance. Hence
$
q_t^{\mathrm{label}}\ge \frac{2}{|\bar{\mathcal X}|}.
$
Again by \eqref{eq:conf-bound-t},
\[
r_t
>
0.9\left(\frac{2}{|\bar{\mathcal X}|}-\frac{0.2}{|\bar{\mathcal X}|}\right)
=
\frac{1.62}{|\bar{\mathcal X}|}
>
\frac{3}{2|\bar{\mathcal X}|}.
\]
Therefore the algorithm takes the first branch and returns
$
\mathcal R_{t+1}
=
[\ell_t+\eta_t-2^{1-m_t},\;b_t].
$
Since
\[
y(x^\star)\ge \ell_t+\eta_t-2^{-m_t}>\ell_t+\eta_t-2^{1-m_t},
\]
we again have \(y(x^\star)\in \mathcal R_{t+1}\).

\medskip
\noindent
\textbf{Case 3:}
$
y(x^\star)\in
[\ell_t+\eta_t-2^{1-m_t},\;\ell_t+\eta_t-2^{-m_t}).
$
If the algorithm takes the first branch, then
$
\mathcal R_{t+1}
=
[\ell_t+\eta_t-2^{1-m_t},\;b_t],
$
so clearly \(y(x^\star)\in \mathcal R_{t+1}\).
If the algorithm takes the second branch, then
$
\mathcal R_{t+1}
=
[a_t,\;\ell_t+\eta_t-2^{-m_t}],
$
and again \(y(x^\star)\in \mathcal R_{t+1}\).
Thus, in all cases,
$
y(x^\star)\in \mathcal R_{t+1}.
$
Therefore, conditioned on \eqref{eq:conf-bound-t}, the returned interval always contains the optimal value. Since \eqref{eq:conf-bound-t} holds with probability at least \(1-\delta_t\), the correctness claim follows.

Next we bound the length of the returned interval. If the first branch is taken, then
$
\mathcal R_{t+1}
=
[\ell_t+\eta_t-2^{1-m_t},\,b_t],
$
and hence
\[
|\mathcal R_{t+1}|
=
b_t-(\ell_t+\eta_t-2^{1-m_t}).
\]
Using
\[
\ell_t=\frac{a_t+b_t}{2},
\qquad
\eta_t=\frac{b_t-a_t}{4},
\]
we obtain
\[
b_t-(\ell_t+\eta_t)
=
\frac{b_t-a_t}{4}
=
\eta_t.
\]
Therefore
\[
|\mathcal R_{t+1}|=\eta_t+2^{1-m_t}.
\]
By the definition of \(m_t\),
\[
2^{-m_t}\le \frac{\eta_t}{2},
\qquad\text{hence}\qquad
2^{1-m_t}\le \eta_t.
\]
Thus
\[
|\mathcal R_{t+1}|
\le
2\eta_t
=
\frac{1}{2}(b_t-a_t)
=
\frac{1}{2}|\mathcal R_t|.
\]

If the second branch is taken, then
$
\mathcal R_{t+1}
=
[a_t,\;\ell_t+\eta_t-2^{-m_t}],
$
so
\[
|\mathcal R_{t+1}|
=
\ell_t+\eta_t-2^{-m_t}-a_t.
\]
Again using the definitions of \(\ell_t\) and \(\eta_t\),
\[
\ell_t+\eta_t-a_t
=
\frac{3}{4}(b_t-a_t)
=
\frac{3}{4}|\mathcal R_t|.
\]
Hence
\[
|\mathcal R_{t+1}|
=
\frac{3}{4}|\mathcal R_t|-2^{-m_t}.
\]
By the definition of \(m_t\),
\[
2^{-m_t}\ge \frac{\eta_t}{4}
=
\frac{1}{16}|\mathcal R_t|.
\]
Therefore
\[
|\mathcal R_{t+1}|
\le
\frac{3}{4}|\mathcal R_t|-\frac{1}{16}|\mathcal R_t|
=
\frac{11}{16}|\mathcal R_t|.
\]

Combining the two branches gives
\[
|\mathcal R_{t+1}|\le \frac{11}{16}|\mathcal R_t|.
\]

Finally, by Lemma~\ref{lemma: complexity A}, the query complexity of constructing \(\mathcal A_t\) is
\[
T_{\mathcal A_t}
=
O\!\left(
\frac{1}{\eta_t}\log\frac{1}{\alpha_t}
\right),
\]
where
\[
\alpha_t=\frac{0.01\delta_t^2}{4|\bar{\mathcal X}|^{3}}.
\]

Moreover, since the auxiliary solution is always labeled, the ideal labeled probability satisfies
\[
q_t^{\mathrm{label}}\ge \frac{1}{|\bar{\mathcal X}|}.
\]
Using \eqref{eq:prob-diff-t}, we obtain
\[
p_t^{\mathrm{label}}
\ge
q_t^{\mathrm{label}}-\frac{0.2}{|\bar{\mathcal X}|}
\ge
\frac{0.8}{|\bar{\mathcal X}|}.
\]
Hence
\[
\frac{1}{\sqrt{p_t^{\mathrm{label}}}}
=
O\!\left(\sqrt{|\bar{\mathcal X}|}\right).
\]

Applying Lemma~\ref{lemma: estimate-amplify} to the estimate call at iteration \(t\), its query complexity is
\[
O\!\left(
\frac{T_{\mathcal A_t}}{\sqrt{p_t^{\mathrm{label}}}}
\log\frac{1}{\delta_t}
\right)
=
O\!\left(
\frac{\sqrt{|\bar{\mathcal X}|}}{\eta_t}\,
\operatorname{polylog}\!\left(\frac{|\bar{\mathcal X}|}{\delta_t}\right)
\right).
\]
Since \(|\bar{\mathcal X}|=|\mathcal X|+1\), this is
\[
O\!\left(
\frac{\sqrt{|\mathcal X|+1}}{\eta_t}\,
\operatorname{polylog}\!\left(\frac{|\mathcal X|+1}{\delta_t}\right)
\right).
\]
This proves the lemma.

\section{Proof of Lemma \ref{lem:confidence-interval}}

Let $\mathcal{R}_t$ denote the region at the beginning of iteration $t$, with $\mathcal{R}_0=[0,1]$. For each \(t\ge 0\), let \(E_t\) be the event that iteration \(t\) preserves the optimal value, namely,
\[
y(x^\star)\in \mathcal R_t \;\Longrightarrow\; y(x^\star)\in \mathcal R_{t+1}.
\]
Consider the call \(\textsc{OptimalRegion}(\mathcal Q,\epsilon,\delta/2)\) in Algorithm~\ref{alg:grover}. For this call, the risk level passed to Algorithm~\ref{alg:confidence_interval} is \(\delta/2\). Hence, at iteration \(t\), the algorithm uses
\[
\delta_t=\frac{\delta/2}{2^{t+1}}=\frac{\delta}{2^{t+2}}.
\]
By Lemma~\ref{lem:shrink-single},
$
\mathbb{P}(E_t^c)\le \delta_t.
$
Therefore, by the union bound,
\[
\mathbb P\!\left(\bigcap_{t\ge 0} E_t\right)
\ge
1-\sum_{t\ge 0}\mathbb P(E_t^c)
\ge
1-\sum_{t\ge 0}\delta_t
=
1-\sum_{t\ge 0}\frac{\delta}{2^{t+2}}
= 1-\frac{\delta}{2}.
\]
On the event \(\bigcap_{t\ge 0}E_t\), since \(y(x^\star)\in \mathcal R_0=[0,1]\), a simple induction on \(t\) shows that
\[
y(x^\star)\in \mathcal R_t
\qquad\text{for all } t\ge 0.
\]

Moreover, by Lemma~\ref{lem:shrink-single},
\[
|\mathcal R_{t+1}|\le \frac{11}{16}|\mathcal R_t|.
\]

Since \(|\mathcal R_0|=1\), it follows that
\[
|\mathcal R_t|\le \left(\frac{11}{16}\right)^t
\qquad\text{for all } t\ge 0.
\]

Hence, the algorithm must terminate once
\[
\left(\frac{11}{16}\right)^t<\frac{\epsilon}{2},
\]
that is, after at most
\[
T=
\left\lceil
\log_{16/11}\frac{2}{\epsilon}
\right\rceil
\]
iterations. Therefore, the returned interval \(\mathcal R\) satisfies
\[
|\mathcal R|<\frac{\epsilon}{2},
\]
and, on the event \(\bigcap_{t\ge 0}E_t\), still contains \(y(x^\star)\). This proves the correctness claim.

For the query complexity, consider iteration \(t<T\). Before termination we have
\[
|\mathcal R_t|\ge \frac{\epsilon}{2},
\qquad
\eta_t=\frac{|\mathcal R_t|}{4}.
\]
By Lemma~\ref{lem:shrink-single}, the query complexity of iteration \(t\) is
\[
O\!\left(
\frac{\sqrt{|\mathcal X|}}{\eta_t}\,
\operatorname{polylog}\!\left(\frac{|\mathcal X|}{\delta_t}\right)
\right)
=
O\!\left(
\frac{\sqrt{|\mathcal X|}}{|\mathcal R_t|}\,
\operatorname{polylog}\!\left(\frac{|\mathcal X|}{\delta_t}\right)
\right).
\]
Now, since $\delta_t=\frac{\delta}{2^{t+2}},$ we have
\[
\operatorname{polylog}\!\left(\frac{|\mathcal X|}{\delta_t}\right)
=
\operatorname{polylog}\!\left(\frac{2^{t+2}|\mathcal X|}{\delta}\right).
\]
Moreover, because \(t\le T\) and \(T=O(\log(1/\epsilon))\), we have \(2^t=O(1/\epsilon)\). Hence
\[
\operatorname{polylog}\!\left(\frac{2^{t+2}|\mathcal X|}{\delta}\right)
=
\operatorname{polylog}\!\left(\frac{|\mathcal X|}{\delta\epsilon}\right).
\]
Therefore, the query complexity of iteration \(t\) is
\[
O\!\left(
\frac{\sqrt{|\mathcal X|}}{|\mathcal R_t|}\,
\operatorname{polylog}\!\left(\frac{|\mathcal X|}{\delta\epsilon}\right)
\right).
\]

Since the interval lengths shrink geometrically, the total cost is dominated by the final iteration. At termination we have \(|\mathcal R_t|=\Theta(\epsilon)\), so summing over all iterations gives
\[
O\!\left(
\frac{\sqrt{|\mathcal X|}}{\epsilon}\,
\operatorname{polylog}\!\left(\frac{|\mathcal X|}{\delta\epsilon}\right)
\right).
\]
This proves the claimed complexity bound.

\section{Proof of Theorem \ref{thm:gas}}
Algorithm~\ref{alg:grover} first calls
$
\mathcal{R}=\textsc{OptimalRegion}(\mathcal Q,\epsilon,\delta/2)
$.
By Lemma~\ref{lem:confidence-interval}, with probability at least \(1-\delta/2\), the returned interval \(\mathcal R=[a,b]\) satisfies $y(x^\star)\in\mathcal{R}$ and $|\mathcal{R}|<\frac{\epsilon}{2}.$
Let \(E_1\) denote this event. In the following, we condition on \(E_1\).

Algorithm~\ref{alg:grover} then sets
\[
\eta=\frac{b-a}{4},
\qquad
\ell=a-\frac{3}{4}\eta,
\]
and constructs the flag algorithm
$
\mathcal A_\mathrm{flag}=\mathcal A(\mathcal Q,\ell,\eta,0.1\delta).
$
Since
\[
m=\left\lceil \log_2\frac{1}{2\eta}\right\rceil+2,
\]
we have
\[
\frac{\eta}{4}\le2^{-m}\le \frac{\eta}{2},
\qquad
2^{1-m}\le \eta.
\]

Observe that
\[
\ell+\eta-2^{-m}
=
a-\frac{3}{4}\eta+\eta-2^{-m}
=
a+\frac{1}{4}\eta-2^{-m}.
\]
Using $2^{-m}\ge \eta/4$, we obtain
\[
\ell+\eta-2^{-m}\le a+\frac{1}{4}\eta-\frac{1}{4}\eta
= a.
\]
Because $y(x^\star)\in[a,b]$, we have
\[
y(x^\star)\ge a\ge\ell+\eta-2^{-m}.
\]

Hence there exists \(x\in\mathcal X\) such that
$
y(x)\ge \ell+\eta-2^{-m},
$
and therefore the assumption of Lemma~\ref{lemma: estimate-amplify} is satisfied.

Next we show that every solution \(x\in\mathcal X\) satisfying
$
y(x)\ge \ell+\eta-2^{1-m}
$
is \(\epsilon\)-optimal. 
Using \(2^{1-m}\le \eta\), we get
\[
y(x)\ge \ell
=
a-\frac{3}{4}\eta
=
a-\frac{3}{16}(b-a).
\]
Since \(b-a<\epsilon/2\), it follows that
\[
y(x)>a-\frac{3\epsilon}{32}.
\]
On the other hand, since \(y(x^\star)\in[a,b]\) and \(b-a<\epsilon/2\), we have
\[
a>y(x^\star)-\frac{\epsilon}{2}.
\]
Combining the above inequalities yields
\[
y(x)
>
y(x^\star)-\frac{\epsilon}{2}-\frac{3\epsilon}{32}
=
y(x^\star)-\frac{19\epsilon}{32}
>
y(x^\star)-\epsilon.
\]
Therefore, every solution $x\in\mathcal{X}$ satisfying
$
y(x)\ge \ell+\eta-2^{1-m}
$ is \(\epsilon\)-optimal.

Now apply
\[
\hat x^\star=\textsc{Amplify}(\mathcal A_{\mathrm{flag}},\delta/2).
\]
By Lemma~\ref{lemma: estimate-amplify}, conditioned on \(E_1\), with probability at least \(1-\delta/2\), the amplification procedure outputs some
$
\hat x^\star
$ such that $y(\hat{x}^\star)\ge \ell+\eta-2^{1-m}$.
By the argument above, this solution is \(\epsilon\)-optimal. Hence, conditioned on \(E_1\),
\[
y(\hat x^\star)\ge y(x^\star)-\epsilon
\]
with probability at least \(1-\delta/2\).

Define the event
\begin{equation*}
    E_2 = \left\{y(\hat{x}^\star)\ge \ell+\eta-2^{1-m}\right\}.
\end{equation*}
Then
\[
\mathbb{P}(E_1)\ge 1-\frac{\delta}{2},
\qquad
\mathbb{P}(E_2\mid E_1)\ge 1-\frac{\delta}{2}.
\]
Therefore,
\[
\mathbb{P}\!\bigl(y(\hat x^\star)\ge y(x^\star)-\epsilon\bigr)
\ge
\mathbb{P}(E_1\cap E_2)
=
\mathbb{P}(E_1)\mathbb{P}(E_2\mid E_1)
\ge
\left(1-\frac{\delta}{2}\right)^2
\ge 1-\delta.
\]
This proves the correctness guarantee.

It remains to bound the query complexity.  
By Lemma~\ref{lem:confidence-interval}, the call to
$
\textsc{OptimalRegion}(\mathcal Q,\epsilon,\delta/2)
$
uses
\[
O\!\left(
\frac{\sqrt{|\mathcal X|}}{\epsilon}\,
\operatorname{polylog}\!\left(\frac{|\mathcal X|}{\delta\epsilon}\right)
\right)
\]
queries.

For the amplification step, Lemma~\ref{lemma: estimate-amplify} gives query complexity
\[
O\!\left(
\frac{T_{\mathcal A}}{\sqrt{p_{\mathrm{label}}}}
\log\frac{1}{\delta}
\right),
\]
where \(T_{\mathcal A}\) is the query complexity of the flag algorithm \(\mathcal A_{\mathrm{flag}}\). 

Under the present condition that there exists \(x\in\mathcal X\) such that
$
y(x)\ge \ell+\eta-2^{-m},
$
the proof of Lemma~\ref{lemma: estimate-amplify} implies that
\[
\frac{1}{\sqrt{p_{\mathrm{label}}}}=O(\sqrt{|\mathcal X|}).
\]
Moreover, by the complexity guarantee of the flag construction and the fact that
\[
\eta=\frac{b-a}{4}=O(\epsilon),
\]
we have
\[
T_{\mathcal A}
=
O\!\left(
\frac{1}{\epsilon}\,
\operatorname{polylog}\!\left(\frac{|\mathcal X|}{\delta}\right)
\right).
\]
Therefore the amplification step uses
\[
O\!\left(
\frac{\sqrt{|\mathcal X|}}{\epsilon}\,
\operatorname{polylog}\!\left(\frac{|\mathcal X|}{\delta}\right)
\right)
\]
queries.
Adding the costs of the two parts, Algorithm~\ref{alg:grover} uses
\[
O\!\left(
\frac{\sqrt{|\mathcal X|}}{\epsilon}\,
\operatorname{polylog}\!\left(\frac{|\mathcal X|}{\delta\epsilon}\right)
\right)
\]
queries in total. Treating \(\epsilon\) as a fixed constant, this is
$
\tilde O(\sqrt{|\mathcal X|}).
$
This completes the proof.
\end{APPENDIX}
%
%


\bibliographystyle{informs2014} 
\bibliography{reference} 

\end{document}